\documentclass{iopart}
\usepackage{epsf}
\begin{document}
\title{Three-dimensional wedge filling in ordered and disordered systems}
\author{M. J. Greenall\dag\footnote[3]{Present address: School of Physics,
The University of Edinburgh, Mayfield Road,
Edinburgh EH9 3JZ, United Kingdom}, A. O. Parry\dag, and J. M. Romero-Enrique\dag\ddag}
\address{\dag\ Department of Mathematics, Imperial College London, 180 Queen's Gate,
London SW7 2BZ, United Kingdom}
\address{\ddag\ Departamento de F\'{\i}sica At\'omica, Molecular y
Nuclear, Area de F\'{\i}sica Te\'orica, Universidad de Sevilla,
Apartado de Correos 1065, 41080 Sevilla, Spain}              
\begin{abstract}
We investigate interfacial structural and fluctuation effects occurring at
continuous filling transitions in 3D wedge geometries. We show that 
fluctuation-induced wedge covariance relations that have been reported
recently for 2D filling and wetting have mean-field or classical analogues that
apply to higher-dimensional systems. Classical wedge covariance emerges from
analysis of filling in shallow wedges based on a simple interfacial 
Hamiltonian model and is supported by detailed numerical investigations of 
filling within a more microscopic Landau-like density functional theory. 
Evidence is presented that classical wedge covariance is also obeyed for 
filling in more acute wedges in the asymptotic critical regime. For
sufficiently short-ranged forces mean-field predictions for the filling
critical exponents and covariance are destroyed by pseudo-one-dimensional
interfacial fluctuations. In this filling fluctuation regime we argue that the 
critical exponents describing the divergence of lengthscales are 
related to values of the interfacial wandering exponent $\zeta(d)$ defined 
for planar interfaces in (bulk) two-dimensional ($d=2$) and three-dimensional
($d=3$) systems. For the interfacial height $l_w\sim 
(\theta-\alpha)^{-\beta_w}$, with $\theta$ the contact angle and 
$\alpha$ the wedge tilt angle, we find $\beta_w=\zeta(2)/2(1-\zeta(3))$. 
For pure systems (thermal disorder) we recover the known result 
$\beta_w=1/4$ predicted by interfacial Hamiltonian studies
whilst for random-bond disorder we predict the universal critical exponent 
$\beta\approx 0.59$ even in the presence of dispersion forces. We revisit 
the transfer matrix theory of three-dimensional filling based on an 
effective interfacial Hamiltonian model and discuss the interplay between 
breather, tilt and torsional interfacial fluctuations. We show that the 
coupling of the modes allows the problem to be mapped onto a quantum 
mechanical problem as conjectured by previous authors.
The form of the interfacial height probability distribution function
predicted by the transfer matrix approach is
shown to be consistent with scaling and thermodynamic requirements for
distances close to and far from the wedge bottom respectively.
\end{abstract}
\pacs{68.08.Bc, 05.70.Np, 68.35.Md, 68.35.Rh}
\maketitle

\section{Introduction\label{introduction}}

Fluids adsorbed at micropatterned and sculpted solid surfaces may
exhibit novel phase transitions and fluctuation effects compared to wetting 
behaviour at planar, heterogeneous walls \cite{Gau,Rascon,Bruschi}. 
A striking example of the influence of substrate geometry on fluid adsorption 
is provided by a simple wedge geometry \cite{Pomeau,Hauge,Rejmer,Parry1,
Parry2,Parry3,Parry4,Parry5}. At two phase coexistence, a wedge-vapour 
interface is completely filled with liquid provided the contact angle $\theta$
is less than the wedge tilt angle $\alpha$. The phase transition from
microscopic to macroscopic adsorption as $\theta\to \alpha^+$ is referred to as
filling and may be first-order or continuous. The conditions for continuous
wedge (and also conic) filling are less restrictive than for continuous 
(critical) wetting at planar walls \cite{Parry1,Parry2} and give some hope 
that large-scale fluctuation effects associated with interfacial unbinding 
may be observable in the laboratory. Recent effective interfacial
Hamiltonian studies have shown that fluctuation effects at continuous
filling transitions exhibit a number of intriguing features. For
three-dimensional wedge filling transitions critical singularities are believed
to be far stronger than those characteristic of critical wetting transitions
reflecting the anisotropy of soft-mode interfacial fluctuations induced by the 
wedge geometry. In particular for pure systems with sufficiently 
short-ranged forces the mean interfacial height $l_w$ (measured above 
the wedge bottom) and roughness $\xi_{\perp}$ are comparable and diverge 
with the same universal critical exponent $\beta_w=\nu_{\perp}=1/4$ 
\cite{Parry1,Parry2}. These predictions are in very good agreement with 
Monte Carlo simulation studies of filling within more microscopic Ising 
and lattice polymer models \cite{Milchev1,Milchev2}. For two-dimensional 
wedge filling on the other hand fluctuation effects are interesting for a 
different reason. Studies of fluctuation-dominated filling in both ordered 
(pure) \cite{Parry3} and disordered (random-bond) \cite{Parry4} systems 
show that some observables, such as the mid-point height probability 
distribution function, show scaling properties which are \emph{identical} 
with short-ranged critical wetting transitions. The only influence of the 
wedge geometry is to shift the effective value of the contact angle from 
$\theta$ to $\theta-\alpha$ - a feature which has been referred to as 
{\it{wedge covariance}} \cite{Parry5}. This ``hidden symmetry" between 
wetting and filling appears to restrict the allowed values of the critical 
exponents at both 2D filling \emph{and} wetting and leads to new some 
insights into the properties of critical wetting transitions.

The present paper focuses on the structural and fluctuation properties of 
3D wedge filling transitions. We begin with a discussion of wedge covariance 
and illustrate using a simple interfacial model of filling in
shallow wedges that the fluctuation-induced covariance observed for 2D 
systems has a mean-field or classical precursor for wetting and filling 
in systems with short-ranged forces. Thus if $l_{\pi}(\theta)$ denotes 
the mean-field result for the contact angle dependence of the critical 
wetting layer thickness, then in the wedge geometry the corresponding 
mean-field result for the mid-point height at bulk coexistence is
\begin{equation}
l_w(\theta,\alpha)=l_{\pi}(\theta-\alpha)
\label{central1}
\end{equation}
The predictions of the interfacial model are supported by a detailed numerical 
study of filling in a Landau-like density functional model that indicate this
classical covariance is obeyed for both shallow and more acute wedges. The
implications of classical wedge covariance for the structure of
interfacial models of short-ranged filling and wetting are discussed. In the
second part of our paper we turn our attention to fluctuation effects at 3D
wedge filling and present a general scaling argument that relates the 
values of the critical exponents for 3D wedge filling to the wandering 
exponent of planar, two-dimensional and three-dimensional fluid interfaces. 
Thus, for example the critical exponent $\beta_w$ is identified as
\begin{equation}
\beta_w=\frac{\zeta(2)}{2(1-\zeta(3))}
\label{central2}
\end{equation}
where $\zeta(d)$ is the standard wandering exponent for a planar-like interface
in a $d$-dimensional bulk system. Our expressions recover the values quoted 
above for pure systems, corresponding to thermal fluctuations, and also 
allow us to discuss 3D filling in disordered systems. We also discuss the 
cross-over from three-dimensional to two-dimensional wedge filling as one 
lowers the number of dimensions of translational invariance along the wedge. 
The change in the nature of fluctuation effects and critical singularities 
in this wedge compactification process highlights a remarkable numerical
coincidence concerning the value of the wandering exponent for planar
interfaces at the lower marginal dimension for wedge filling. In the final 
part of our study we focus on 3D filling in pure systems and address some 
problems that have been highlighted concerning the transfer-matrix analysis 
of a pseudo-one-dimensional effective Hamiltonian model. We argue that 
problems associated with the choice of the appropriate measure in the
functional integral can be avoided if, in addition to breather-mode 
excitations of the interface, we also allow for tilt and torsional degrees 
of freedom. This more accurate formulation of the theory allows us to 
determine the universal scaling function associated with the density profile 
in the wedge. The form of this function is shown to have the correct 
short-distance and large-distance behaviour dictated by scaling theory and 
macroscopics respectively. 
\section{Background theory\label{background}}

\subsection{Thermodynamics and critical exponents\label{background1}}

\begin{figure}
\begin{center}
\epsfxsize=9cm
\epsfbox{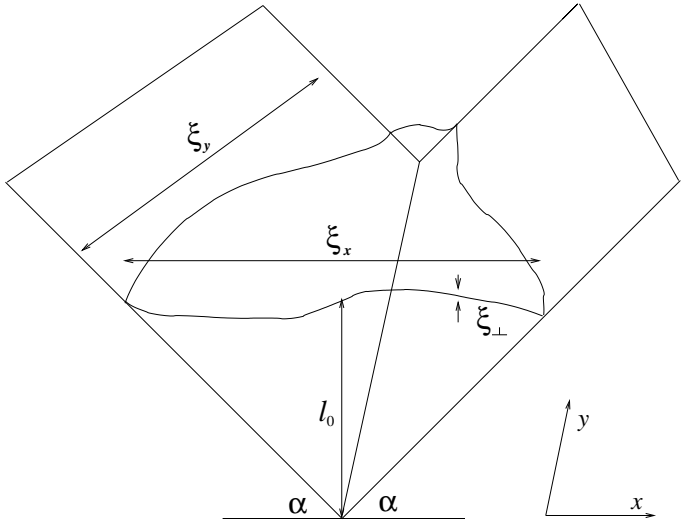}
\end{center}
\caption{Schematic illustration of a typical interfacial configuration
in the 3D wedge geometry and the typical diverging lengthscales at the
filling transition. Note that $l_w=\langle l_0 \rangle$. \label{fig1}}
\end{figure}

Consider a 3D wedge formed from the intersection of two smooth, planar walls
that meet at angles $\alpha$ to the $z=0$ plane forming a wedge with opening 
angle $\pi-2\alpha$. The parallel displacement vector in the $z=0$ plane is
written ${\bf{x}}=(x,y)$ with cartesians $(x,y)$ measuring distances across and
along the wedge respectively. Thus the height of the wall above the $z=0$ 
plane is described by a height function $z_w({\bf{x}})=\tan\alpha \vert x 
\vert$. We suppose that the wedge is in contact with a three-dimensional
bulk vapour at sub-critical temperature $T$ and chemical potential $\mu$ tuned
to bulk two-phase coexistence $\mu=\mu_{sat}(T)^-$.  Provided the contact 
angle $\theta(T)$ of the sessile drop (defined for the planar wall-fluid 
interface) is less than ninety degrees the wedge preferentially adsorbs a 
volume of liquid near its bottom. Partial and complete filling refer to 
situations where the adsorption is microscopic and macroscopic respectively. 
The separatrix between partial and complete filling follows from simple 
thermodynamic arguments first discussed by Concus and Finn \cite{Concus}
some thirty years ago (see also Pomeau \cite{Pomeau} and Hauge \cite{Hauge}). 
Let $V$ denote the accessible volume of  the fluid and $A$, $L$ the area and 
length of the wedge respectively. The total grand potential $\Omega$ contains 
the macroscopic contributions
\begin{equation}
\Omega=-pV+\sigma_{wv}A+f_w L
\label{thermo2}
\end {equation}
where $\sigma_{wv}$ is the surface tension of the wall-vapour interface and
$f_w$ is the excess wedge free energy. Now suppose we are at bulk coexistence 
and imagine that the wedge is filled to a height $l_w$ (see \Fref{fig1}).
Macroscopically the liquid-vapour interface must be flat and hence $f_w$ 
must contain a thermodynamic contribution 
\begin{equation}
f_w=\frac{2\sigma_{lv}(\cos \alpha-\cos\theta)l_w}{\sin\alpha}
\label{thermo3}
\end{equation}
where $\sigma_{lv}$ is the liquid-vapour tension and we have used Young's 
equation. It follows that for $\theta<\alpha$ the free energy can be 
lowered by completely filling the wedge. For $\theta>\alpha$ on the other 
hand the equilibrum value of $l_w$ must be finite and arises from a balance 
between thermodynamics and internal energy (intermolecular forces) or 
entropic (fluctuation) contributions. Thus by either increasing the 
tilt angle at constant $T$ or increasing the temperature at constant angle 
(assuming the usual scenario where the contact angle decreases with 
temperature) one can induce a transition from partial to complete filling 
at a temperature $T_{fill}$ satisfying 
\begin{equation}
\theta(T_{fill})=\alpha
\label{thermo}
\end{equation}
Filling transitions may be either first- or second-order corresponding 
to the discontinuous or continuous divergence of the mean interfacial 
height $l_w$ respectively. Importantly, the conditions for continuous filling 
are less restrictive than for continuous wetting transitions and wedges made 
from walls that exhibit first-order wetting (at some higher temperature 
$T_{wet}$) may still exhibit continuous filling \cite{Parry1,Parry2}. 
The most important diverging lengthscales which characterise the the 
transition are the mean mid-point height $l_w$, the perpendicular correlation 
length $\xi_{\perp}$ at the mid-point defined from the root-mean-square 
variance, and the correlation length $\xi_y$ describing fluctuations along 
the wedge. Correlations across the wedge are described by a lengthscale which 
is trivially related to the mean-height: $\xi_x\sim l_w\cot\alpha$. At bulk
coexistence critical exponents for the divergence of these length scales are
identified according to 
\begin{equation}
l_w\sim t^{-\beta_w}\quad,\quad \xi_{\perp}\sim
t^{-\nu_{\perp}}\quad,\quad \xi_y\sim
t^{-\nu_y}
\label{fillingexponents}
\end{equation}
 where the temperature-like scaling field $t\propto(T_{fill}-T)\propto
(\theta-\alpha)$. The phenomenology is very similar for the case of 2D wedge
filling except there is no analogue of the correlation length $\xi_y$. 

\subsection{3D wedge filling in pure systems\label{background2}}

A suitable starting point for the evaluation of the critical exponents is an
effective interfacial Hamiltonian model based on a collective co-ordinate
$l({\bf{x}})$ measuring the height of the unbinding interface above the $z=0$
plane. At least for shallow tilt angles $\alpha\sim \tan\alpha$, the
appropriate effective Hamiltonian is a natural generalization of the standard
capillary-wave model used to study planar wetting and we write \cite{Rejmer}
\begin{equation}
H[l] = \int d{\bf{x}} \left\{ \frac{\Sigma}{2} \left(\nabla l \right)^2
+W(l-\alpha\vert x \vert) \right\}
\label{Heff}
\end{equation}
where $\Sigma$ is the stiffness coefficient of the unbinding interface which,
for isotropic fluids, we can identify with the surface tension $\sigma_{lv}$. 
The model is not truly microscopic and is only valid for lengthscales much 
larger than the bulk correlation length. The binding potential $W(l)$
describes the influence of intermolecular forces and decays at large distances
as $W(l)\sim-a l^{-p}$. The Hamaker constant $a$ is positive in the
temperature region of interest whilst the value of exponent $p$ depends on the
specific range of the forces with $p=2,3$ for non-retarded and retarded
van der Waals forces respectively. It is also possible to modify the model to
include the effect of quenched disorder but for the moment we concentrate on
pure systems with thermal fluctuations. Minimisation of the capillary-wave-like
model (\ref{Heff}) determines the following mean-field values of the critical
exponents for 3D filling \cite{Parry1} 
\begin{equation} 
\beta_w=\frac{1}{p}\quad,\quad\nu_{\perp}=\frac{1}{4}\quad,\quad 
\nu_y=\frac{1}{2}+\frac{1}{p}
\label{mf}
\end{equation}
Thus fluctuation effects are extremely anisotropic at wedge filling with
$\xi_y\gg \xi_x$ and are dominated by pseudo-one-dimensional local 
translations in the height of the filled region along the wedge 
\cite{Parry1,Parry2}. We refer to these as ``breather-mode" excitations. 
From the values of the above critical exponents it is clear that mean-field 
theory is only valid for intermolecular potentials with $p<4$ for which 
$\xi_{\perp}\ll l_w$. The filling fluctuation (FFL) regime corresponds to 
potentials with $p>4$ and to study this Parry, Rasc\'{o}n and Wood introduced a 
pseudo-one-dimensional wedge Hamiltonian which accounts only for the 
breather-mode excitations \cite{Parry1,Parry2}:
\begin{equation}
H_w[l_0]=\int dy \left\{ \frac{\Sigma l_0}{\alpha} \left(\frac{ dl_0}{dy} 
\right)^2+V(l_0) \right\}
\label{Hw}
\end{equation}
where $l_0(y)\equiv l(0,y)>0$ is the local height of the interface above the
wedge bottom and $l_w=\langle l_0(y)\rangle$. 
The model is considered valid only for small wave-vectors
$k_y\ll k_y^{max}\sim\xi_x^{-1}$. However because the fluctuations at filling 
are strongly anisotropic the relevant scaling combination $\xi_yk_y^{max}$ 
diverges in the scaling limit and the cut-off does not determine universal 
quantities. The most important feature of this effective model is the presence 
of a bending term resisting fluctuations along the wedge which is proportional 
to the local interfacial height. At bulk coexistence the wedge potential 
$V(l_0)$ has the form 
\begin{equation}
V(l_0)=\frac{\Sigma}{\alpha}(\theta^2-\alpha^2)l_0 + C l_0^{1-p}
\label{wedgepotential}
\end{equation}
and for $\theta>\alpha$ has a minimum located at the mean-field value of $l_w$.
Notice that the first term in the wedge binding potential is the small $\alpha$
approximation to the thermodynamic term $f_w$ and is proportional to the 
linear, temperature-like scaling field $t$. The transfer-matrix analysis of the
one-dimensional model will be considered later. Here we point out two related
interpretations of the model which give the desired classification of the 
allowed critical singularities without need for explicit calculations. 
Both arguments will play a role in our later discussions. To begin we 
observe that the bending energy term is invariant under a renormalization 
group rescaling
\begin{equation}
y\to y'=\frac{y}{b}\quad,\quad l_0\to l_0'=\frac{l_0}{b^{\zeta_w}}
\label{simplerescaling1}
\end{equation}
with wedge wandering exponent $\zeta_w=1/3$. Under this transformation the 
linear scaling field $t$ and wedge Hamaker constant $C$ rescale to $t'$ and 
$C'$ with
\begin {equation}
t'=b^{{\frac{4}{3}}}t\quad,\quad C'= b^ {{\frac{4-p}{3}}}C
\label{simplerescaling2}
\end{equation}
Thus as anticipated the intermolecular forces are relevant and irrelevant for
$p<4$ and $p>4$ respectively. The temperature-like scaling field is always
relevant and from its rescaling and the value of $\zeta_w$ we can identify the
values of the exponents in the FFL regime as
\begin{equation} 
\beta_w=\frac{1}{4}\quad,\quad \nu_{\perp}=\frac{1}{4}\quad,\quad 
\nu_y=\frac{3}{4}
\label{FFL3D}
\end{equation}
Equivalently we may say that in the FFL the diverging lengths scale according
to 
\begin{equation}
l_w\sim\xi_{\perp}\sim \xi_y^{\frac{1}{3}}
\label{simplescaling3}
\end{equation}
which shows the dramatic influence the wedge geometry has on the fluid
interfacial properties near the filling phase boundary.
Closely related to this simple rescaling analysis is the construction of an
effective potential. Heuristically we may imagine that the equilibrium
interfacial height $l_w$ follows from minimization of an effective wedge
potential which takes into account the contribution arising from the bending
free energy. Noting the scaling relation between $l_w$, $\xi_{\perp}$ and
$\xi_y$ together with the first term in the wedge Hamiltonian it is natural to
suppose that the effective potential has the form
\begin{equation}
V_{eff}(l_0)=V(l_0)+Dl_0^{-\tau_w}
\label{effV1}
\end{equation}
where $\tau_w={2/{\zeta_w}-3}$ and the constant $D\propto \Sigma$. Thus for the
present case of thermal disorder $\tau_w=3$ and the critical behaviour falls 
into two regimes for $p<4$ and $p>4$ depending on whether intermolecular 
forces or the bending energy is the leading order correction to the 
thermodynamic term. 

It is straightforward to generalise these ideas to a $d$-dimensional
wedge which has translational invariance in $d_y=d-2$ dimensions along the 
wedge. As we shall see consideration of the ``compactifaction" from a three- 
to two-dimensional wedge highlights some interesting properties of the 
wandering exponent for planar interfaces. Breather-mode excitations do not 
lead to large-scale interfacial roughness for $d>4$ and mean-field theory 
is valid. For $d<4$ we find two possible fluctuation regimes corresponding 
to mean-field-like and fluctuation-dominated behaviour. The FFL occurs for 
sufficiently short-ranged potentials with $p>2(1/\zeta_w-1)$ with a 
generalised wedge wandering exponent
\begin{equation}
\zeta_w=\frac{4-d}{3}
\label{zetawd}
\end{equation}
valid for $2<d<4$. The critical exponent for the divergence of the interfacial 
height in the FFL regime follows as
\begin{equation}
\beta_w=\frac{4-d}{2(d-1)}
\label{betawd}
\end{equation}
and recovers the $1/4$ power law for the three-dimensional wedge. Intriguingly
the result (\ref{betawd}) has the correct two-dimensional limit $\beta_w=1$ for
filling with thermal fluctuations. We shall return to this 
later but first focus on the 2D limit in more detail.

\subsection{Fluctuation-induced wedge covariance in 2D\label{background3}}

Interfacial Hamiltonian studies \cite{Parry3,Parry4,Parry5} reveal that the 
fluctuation regimes for 2D filling are different to those for 2D critical 
wetting. Nevertheless there appears to be a profound connection between the 
fluctuation-dominated regimes for each transition. Recall that critical 
wetting refers to the continuous unbinding of the liquid-vapour interface 
(say) from a planar wall as the temperature is increased to the wetting 
temperature $T_{wet}$ at bulk coexistence. The vanishing of the contact 
angle $\theta$ as $T\to T_{wet}$ is accompanied by the divergence of the 
mean interfacial thickness $l_{\pi}$, roughness $\xi_{\perp}$ and transverse 
correlation length $\xi_{\parallel}$. More detailed information concerning 
the local height fluctuations is contained in the one-point interfacial height 
probability distribution function (PDF) $P_{\pi}(l;\theta)$ where, in 
preparation for our discussion of wedge covariance, we have written the PDF 
in terms of the contact angle. Wetting transitions are characterised by 
highly anisotropic fluctuations arising from the random-walk or 
capillary-wave-like motion of the interface which are quantified by the 
scaling relation $\xi_{\perp}\sim \xi_{\parallel}^{\zeta(d)}$ with 
$\zeta(d)$ the wandering exponent for a free interface (see later). At 
bulk coexistence the critical exponents for critical wetting describing 
the asymptotic divergence of these lengthscales are identified from
\begin{equation}
l_{\pi}\sim (T_{wet}-T)^{-\beta_s}\ ,\ \xi_{\perp}\sim
(T_{wet}-T)^{-\nu_{\perp}}\ ,\ \xi_{\parallel}\sim
(T_{wet}-T)^{-\nu_{\parallel}}
\label{wettingexponents}
\end{equation}
Generically, fluctuation effects at critical wetting fall into three regimes
depending by the range of the intermolecular forces \cite{Forgacs,Lipowsky1,
Fisher,Lipowsky2,Lipowsky3}. These are mean-field (MF), weak-fluctuation 
(WFL) and strong-fluctuation (SFL) scaling regimes. The SFL regime represents 
the universality class of critical wetting with sufficiently short-ranged 
forces $p>2(1/\zeta(2)-1)$ and is characterised by large fluctuations 
$l_{\pi}\sim\xi_{\perp}$ and universal critical exponents. In two 
dimensions rather general random-walk-based arguments predict that the 
critical exponents in the SFL regime are explicitly related to the value 
of the wandering exponent. In particular for the interfacial height it is 
believed that \cite{Fisher,Lipowsky2}
\begin{equation}
\beta_s=\frac{\zeta(2)}{1-\zeta(2)} 
\label{SFL}
\end{equation}
which is certainly obeyed for pure ($\zeta(2)=1/2$) and random-bond
systems ($\zeta(2)=2/3$).

For 2D filling on the other hand interfacial Hamiltonian studies show 
there are only two fluctuation classes corresponding to mean-field and
fluctuation-dominated regimes for which $l_w\gg \xi_{\perp}$ and
$l_w\sim\xi_{\perp}$ respectively \cite{Parry3,Parry4,Parry5}. In general 
the critical exponents describing mean-field filling and mean-field critical 
wetting transitions are unrelated. Also the filling fluctuation (FFL) regime, 
representing the universality class of systems with short-ranged
forces, is broader than the SFL regime for critical wetting and occurs for 
$p>(1/\zeta(2)-1)$. Simple scaling arguments indicate that within the 
FFL regime the interfacial height diverges with a critical exponent
\begin{equation}
\beta_w=\frac{\zeta(2)}{1-\zeta(2)} 
\label{FFL2D}
\end{equation}
which is in agreement with model calculations in pure \cite{Parry3}
and impure \cite{Parry4} systems. \emph{Wedge covariance} \cite{Parry5}
refers to the ``empirical" observation emerging from effective interfacial 
Hamiltonian and Ising model studies that the equality of
the critical exponents also extends to the full scaling properties of the
respective one-point PDF's. In brief, wedge covariance states that for systems with
short-ranged forces, at bulk coexistence and in the scaling limit, the 
interfacial height PDF at the wedge mid-point $P_w(l;\theta,\alpha)$ is 
identical to the analogous PDF for strong-fluctuation regime critical 
wetting at a planar wall-fluid interface but with an effective shifted
contact angle $\theta\to\theta-\alpha$ with $\alpha$ the wedge tilt angle. 
Thus if $P_{\pi}(l;\theta)$ denotes the critical wetting interfacial height 
PDF written in terms of the contact angle, wedge covariance implies
\begin{equation}
P_w(l;\theta,\alpha)=P_{\pi}(l;\theta-\alpha)
\label{Cov1}
\end{equation}
which is valid in the critical regime $\theta-\alpha\to 0$. From this it
follows that the equilibrium mid-point height near a fluctuation-dominated
filling transition satisfies
\begin{equation}
l_w(\theta,\alpha)=l_{\pi}(\theta-\alpha)  
\label{Cov2}
\end{equation}
where, in an obvious notation, $l_{\pi}(\theta)$ denotes the SFL 
regime critical wetting film thickness expressed in terms of the contact angle.
Interestingly, the covariance  (\ref{Cov1}) necessarily leads to the
exponent identifications (\ref{SFL},\ref{FFL2D}) whilst a third relation 
between the FFL wedge free energy and SFL point tension $\tau(\theta)$ leads 
to the new result for wetting \cite{Parry5}
\begin{equation}
l_{\pi}(\theta)=-\frac{\tau'(\theta)}{2\sigma_{lv}}
\label{tau}
\end{equation}
This expression leads directly to the identification of
the critical singularity associated with the point tension consistent with 
exact Ising model calculations \cite{ALU} and a more general conjecture 
due to Indekeu and Robledo \cite{Indekeu}. We emphasise that the 
above covariance relations, originally  noted from interfacial Hamiltonian 
studies of filling in shallow wedges (in pure and impure systems), appear to 
be rather robust. For pure systems they are also obeyed by a drumhead 
interfacial model of filling in more acute wedges \cite{Abraham1} and, most 
importantly, are consistent with exact \cite{Abraham2} and numerical 
studies \cite{Albano} of filling in the square lattice Ising model with tilt 
angle $\alpha=\pi/4$. 

\section{Classical Wedge Covariance for filling in shallow wedges
\label{shallow}}
 
The standard fluctuation theory of wetting at planar wall-fluid interfaces 
is based on analysis of effective interfacial Hamiltonian models which describe
the fluctuations of a collective co-ordinate $l({\bf{x}})$  measuring the 
local height of the interface from the wall. This coarse-grained description 
is valid at lengthscales much bigger than the bulk correlation 
length and, in its simplest form, is written
\begin{equation}
H_{\pi}[l] = \int d{\bf{x}} \left\{ \frac{\Sigma}{2} \left(\nabla l \right)^2
+W(l) \right\}
\label{Heffplanar}
\end{equation}
where, as described earlier, $\Sigma$ is the stiffness coefficient of the 
unbinding interface and $W(l)$ is the binding potential which, in general, 
accounts for the direct influence of intermolecular forces. We shall 
focus on isotropic fluid interfaces for which $\Sigma$ can be identified 
with the surface tension $\sigma_{lv}$.

For systems with short-ranged forces the binding potential describes not so
much the range of the intermolecular potentials but the decay of perturbations
in the local microscopic order parameter (density, magnetization) and is
usually taken to have the form, at bulk coexistence,
\begin{equation}
W(l)=-a e^{-\kappa l}+b e^{-2\kappa l}
\label{Binding1}
\end{equation}
where $\kappa$ is the inverse bulk correlation length of the wetting phase. 
The temperature dependence of the coefficients $a,b$ is crucial
at mean-field level. The leading order term vanishes at the mean-field 
wetting temperature so that $a\propto (T_{wet}^{MF}-T)$ whilst $b$ must be 
positive finite at $T_{wet}$ to ensure stability and is usually taken to be 
constant. The structure of the binding potential can be inferred by 
comparison with more microscopic Landau-like mean-field theories. More 
formally it can derived from a constrained fluctuation sum as in the 
approach of Fisher and Jin which also leads to the presence of a
position-dependent stiffness-coefficient \cite{FisherJin,Jin}.

At mean-field level the equilibrium thickness $l_{\pi}$ of the 
interface follows from minimization of $W(l)$ whilst the contact angle
$\theta$ can be identified from $\Sigma \theta^2/2=-W(l_{\pi})$. Thus for
systems with short-ranged forces the mean-field results are
\begin{equation}
\kappa l_{\pi}= \ln 2b/a\quad,\quad a=\sqrt{2\Sigma b} \theta
\label{MFl}
\end{equation}
corresponding to critical exponents $\beta_s=0(\ln)$ and $\alpha_s=0$. Written
in terms of the contact angle the equilibrium film thickness of the critical
wetting layer is therefore
\begin{equation}
\kappa l_{\pi}(\theta)= -\ln \sqrt{\frac{\Sigma}{2b}}\theta
\label{MF2}
\end{equation}
which will be our intial point of reference for the wedge calculation.

For shallow wedges corresponding to tilt angles $\alpha\sim\tan\alpha$ 
the natural generalization of the planar model is the effective 
Hamiltonian (\ref{Heff}) which simply assumes the interfacial 
interaction with the wall is controlled by the local relative height 
variable $\tilde l=l-\alpha \vert x\vert$. Note that the mean-field
analysis is essentially independent of dimension since it assumes 
translational invariance along the wedge. The Hamiltonian is minimised 
subject to the appropriate boundary condition $\tilde l\to l_{\pi}$ as 
$\vert x\vert\to\infty$ and yields the Euler-Lagrange equation
\begin{equation}
\Sigma l''(x)=W'(l-\alpha\vert x\vert)
\label{EL}
\end{equation}
It is convenient to write this in terms of the local relative height $\tilde
l=l-\alpha\vert x\vert$ which, making analogy with classical mechanics, 
exploits the local Galilean invariance of the Euler-Lagrange equation. 
On integration one obtains the ``energy" equation 
\begin{equation}
\frac{\Sigma}{2} (\vert l'(x)\vert-\alpha)^2=\Delta W(l-\alpha\vert x\vert)
\label{EL2}
\end{equation}
where $\Delta W(l)=W(l)-W(l_{\pi})$. Thus at the wedge mid-point one finds
the simple expression for the local height of the filling film
\begin{equation}
\frac{\Sigma}{2} \alpha^2=\Delta W(l_w)
\label{MFlw}
\end{equation}
As pointed out by Rejmer \etal \cite{Rejmer} this equation has a very elegant 
graphical interpretation which demonstrates that for quite arbitrary choices 
of binding potential the wedge undergoes a filling transition when $\theta(T)
=\alpha$, in precise accord with thermodynamic arguments. Now consider
the specific case of short-ranged forces with the binding potential
(\ref{Binding1}). Substitution determines the mid-point height as
\begin{equation}
\kappa l_w(\theta,\alpha)= -\ln \sqrt{\frac{\Sigma}{2b}}(\theta-\alpha)
\label{lw}
\end{equation}
This not only identifies the logarithmic divergence of the film thickness at
the filling transition, $\beta_f=0(\ln)$, but also reveals that the mean-field 
theory shows a ``classical" analogue of wedge covariance observed in the 2D 
calculations:
\begin{equation}
l_w(\theta,\alpha)=l_{\pi}(\theta-\alpha)
\label{MFcov1}
\end{equation}
In other words, for systems with short-ranged forces, the influence of the
wedge geometry as manifest in the mid-point height is to shift the effective 
value of the contact angle. A similar property extends to the whole equilibrium
profile which satisfies
\begin{equation}
l(x)-\alpha\vert x\vert=l_{\pi}(\theta-\alpha +\vert l'(x)\vert)
\label{Mfprofile}
\end{equation}
and smoothly interpolates from $l_{\pi}(\theta-\alpha)$ to 
$l_{\pi}(\theta)$ as $\vert x\vert$ increases. Profile covariance
indicates that the height dependence of the meniscus contains information
about the contact angle dependence of the planar wetting film $l_{\pi}(\theta)$
and vice versa.

Classical wedge covariance is also manifest in the Gaussian fluctuations 
about the mean-field solutions. This is most easily interpretated for the 
2D wedge. Consider for example the mean-field expression for the connected 
height-height correlation function
\begin{equation}
S_w(0,x)=\langle (l(0)-\langle l(0)\rangle)(\tilde l(x)-\langle \tilde 
l(x)\rangle)\rangle
\label{Sdef}
\end{equation}
and note that $S_w(0,0)=\xi_{\perp}^2$ identifies the mid-point roughness. The
correlation function can be easily obtained from solution to the appropriate 
Ornstein-Zernike-like equation yielding
\begin{equation}
S_w(0,x)= \frac{\vert \tilde l'(x)\vert}{2 W'(l_w)}
\label{Sgen}
\end{equation}
where, for convenience, we have set $k_B T=1$. For systems with short-ranged 
forces the explicit result for the mean-field roughness at the wedge 
mid-point satisfies
\begin{equation}
\xi_{\perp}(\theta,\alpha)^2= \frac{1}{2\kappa\Sigma (\theta-\alpha)}
\label{Swshort}
\end{equation}     
This can be compared with the result pertinent to critical wetting at the 
planar wall-fluid interface which can be obtained by simply setting 
$\alpha= 0$,
\begin{equation}
\xi_{\perp}(\theta)^2= \frac{1}{2\kappa\Sigma \theta}
\label{Spi}
\end{equation}
 implying the classical wedge covariance relation
\begin{equation}
\xi_{\perp}(\theta,\alpha)=\xi_{\perp}(\theta-\alpha)
\label{MFcov3}
\end{equation}
for Gaussian fluctuations about the mean-field solution.

For higher-dimensional wedges the actual mid-point roughness $\xi_{\perp}$
is no longer covariant because of the breather-mode excitations along the
wedge. For such dimensions the quantity analogous to the roughness which shows
covariance is simply the zeroth moment of the the height-height correlation
function 
\begin{equation}
 S_w(0,x)= \int dy_{12}\langle (l(0)-\langle l(0)\rangle)(\tilde l({\bf{x}})-
\langle \tilde l({\bf{x}})\rangle)\rangle
\label{S3d}
\end{equation}
where $y_{12}$ is the relative separation of the two points along the wedge.
This satisfies the same expression (\ref{Sgen}) quoted above \cite{Parry1}
and is therefore covariant. 

We emphasise that classical covariance is \emph{not} a general feature of the
mean-field theory of filling and critical wetting. Within a more general
description based on a binding potential
\begin{equation}
W(l)=-al^{-p}+bl^{-q}
\label{Wlong}
\end{equation}
the filling and critical wetting exponents are distinct:
\begin{equation}
\beta_w=\frac{1}{p}\quad,\quad \beta_s=\frac{1}{q-p}
\label{betabeta}
\end{equation}
This automatically rules out the possibility of (classical) wedge covariance 
for binding potentials of the form (\ref{Wlong}). Nevertheless the results
presented above showing wedge covariance for short-ranged forces do generalise
to the class of potentials
\begin{equation}
W(l)=-a\omega(l)+b\omega(l)^2
\label{omega}
\end{equation}
where $\omega(l)$ corresponds to an arbitrary choice of monotonically decaying
function. Choosing $\omega=e^{-\kappa l}$ we obtain the usual short-ranged
binding potential whilst setting $\omega=l^{-p}$ one obtains a binding
potential describing a particular type of multicritical wetting transition with
$q=2p$. Repeating the analysis above we obtain for the planar critical wetting
and mid-point wedge filling interfacial heights
\begin{equation}
\omega(l_{\pi})=\sqrt{\frac{\Sigma}{2b}}\theta\quad,\quad \omega(l_w)=\sqrt
{\frac{\Sigma}{2b}}(\theta-\alpha),
\label{GenMFcov}
\end{equation}
which immediately implies the wedge-covariant relation (\ref{MFcov1}). A little
more algebra shows that the roughnesses also obey covariance implying that the
mean-field PDF's are identical provided we map $\theta\to \theta-\alpha$. 
We refer to the class of potentials (\ref{omega}) as classical wedge 
covariant binding potentials. Of course, only for the short-ranged case 
$\omega=e^{-\kappa l}$ do we anticipate that such potentials have any 
physical significance. Nevertheless they do point to an important feature 
of the present analysis. It is easy to demonstrate that the wedge 
covariant potentials (\ref{omega}) all describe planar critical wetting 
transitions with vanishing specific heat exponent $\alpha_s=0$. This 
mirrors precisely the situation for the fluctuation-induced non-classical 
covariance since for $\zeta\ge 1/2$ the value of the specific heat exponent 
in the SFL regime is $\alpha_s=0$. This covers both cases, $\zeta=1/2$ and 
$\zeta=2/3$ in 2D, where non-classical covariance is known to occur and 
it is tempting to speculate that the vanishing of the specific heat 
exponent either at mean-field level or beyond plays a key role
for classical and non-classical covariance respectively.

The prediction of classical wedge covariance shows that this hidden
relation between filling and wetting for special types of forces is not
necessarily a fluctuation-induced phenomenon. Having said that we feel care 
should be taken in saying that the classical covariance is the underlying 
origin of non-classical covariance. Whilst they are certainly related it 
is far from obvious why the covariant relation (\ref{MFcov1}) 
obeyed for binding potentials describing mean-field critical wetting with 
$\alpha_s=0$ should remain unaltered in the presence of large-scale 
fluctuation effects which renormalise the value of the exponent $\beta_s$ 
whilst leaving $\alpha_s$ unchanged. In view of this we treat classical 
covariance as a prediction which should be tested in more
microscopic models of wetting and filling. Since the binding potential
(\ref{Binding1}) is believed to describe wetting in systems with short-ranged
forces we turn attention to numerical studies of filling based on a
Landau-like density functional model. This will also allow to test any
limitations arising from the shallow wedge approximation implicit in the
interfacial model.

\section{Wedge Filling within Landau Theory\label{landau}}
 
For our Landau theory study we resort to a magnetic terminology
rather than the fluids-based one considered earlier. At mean-field level the 
equilibrium order parameter $m({\bf{r}})$ is translationally invariant 
along the wedge so we can restrict ourselves to magnetization profiles 
in a two-dimensional space ${\bf{r}}=(x,z)$ with $x$ the co-ordinate 
across the wedge. The free-energy functional for the infinite wedge that 
we wish to minimise is
\begin{equation}
F[m]=\int_{V}d{\bf{r}}\left\{\frac{1}{2}(\nabla m)^2
-\frac{t}{2}m^2+\frac{u}{4}m^4-hm\right\}
\label{LandauF}
\end{equation}
where the volume of integration is restricted to $z\ge \tan\alpha \vert x
\vert$ for every $x$. The parameter $t$ measures the deviation from the bulk
critical temperature (which is always finite) whilst $u>0$ for stability. The
bulk field $h=0^-$ so that the bulk magnetization is negative. The 
temperature dependence of the equilibrium profiles can be
eliminated by measuring the magnetization in units of the bulk spontaneous
magnetization $m_0(t)=\sqrt{t/u}$ and need not be specified further. Rather 
than use a local surface field and enhancement parameter we use fixed boundary
conditions which set the surface magnetization to a positive value 
$m(x,\tan\alpha \vert x\vert)=m_1$ for all $x$. This is equivalent to the 
infinite surface enhancement limit in the model of Nakanishi and Fisher 
\cite{Nakanishi} and ensures that the wetting transition pertinent to the 
planar wall-down spin interface is always second-order. In the planar limit 
$\alpha=0$ the model can be solved analytically and exhibits a critical 
wetting transition when the surface magnetization
\begin{equation}
m_1^{wet}=m_0
\label{Landauwetting}
\end{equation} 
which allows us to induce wetting (and filling in the wedge) by either 
increasing $m_1$ at fixed $t$ or vary $t$ at fixed $m_1$. We have chosen to 
vary $m_1$ at fixed temperature since this keeps the bulk magnetization 
and correlation length $\xi_b=1/\kappa=(2t)^{-1/2}$ fixed. The contact 
angle within this model can be calculated analytically
\begin{equation}
\cos\theta=\frac{3m_1}{2m_0}\left(1-\frac{m_1^2}{3m_0^2}\right) 
\label{Landautheta}
\end{equation}
so that near the wetting transition $\theta\propto (m_0-m_1)/m_0$ where
 $(m_0-m_1)/m_0$ may be regarded as the temperature-like linear scaling field. 
Thus in the wedge geometry it is straightfoward to convert the numerically-determined value of the surface magnetization at filling phase boundary 
$m_1^{fill}$ into a contact angle $\theta$. We also remark that near the
filling transition the scaling field $\theta-\alpha$ is equivalent to
$(m_1^{fill}-m_1)/m_0$.
\begin{figure}
\epsfxsize=9cm
\begin{center}
\epsfbox{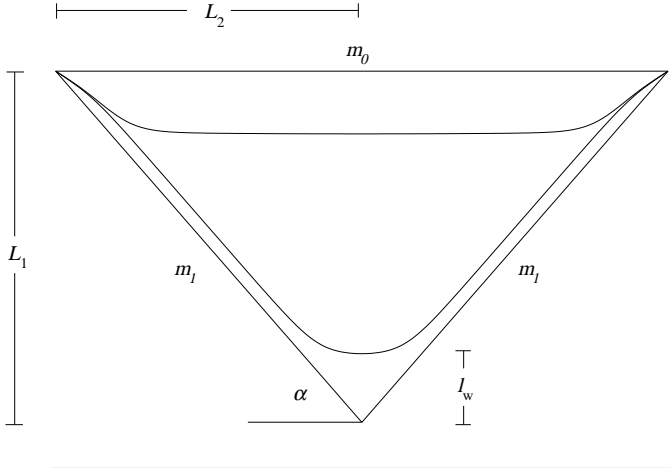}
\end{center}
\caption{Typical capped wedge geometry used for the Landau numerical 
calculations. The magnetisation has a fixed value $m_1$ at the wedge
boundaries and the bulk value $m_0$ at $z=L_1$. Here, $\alpha=45^\circ$ and
$L_1=L_2\approx 30\xi_b$. Two solutions corresponding to either side
of a filling transition are shown: the lower interface for $m_1/m_0=0.5$ and
the upper interface for $m_1/m_0=0.55$.\label{fig2}}
\end{figure}

\begin{figure}
\epsfxsize=9cm
\begin{center}
\epsfbox{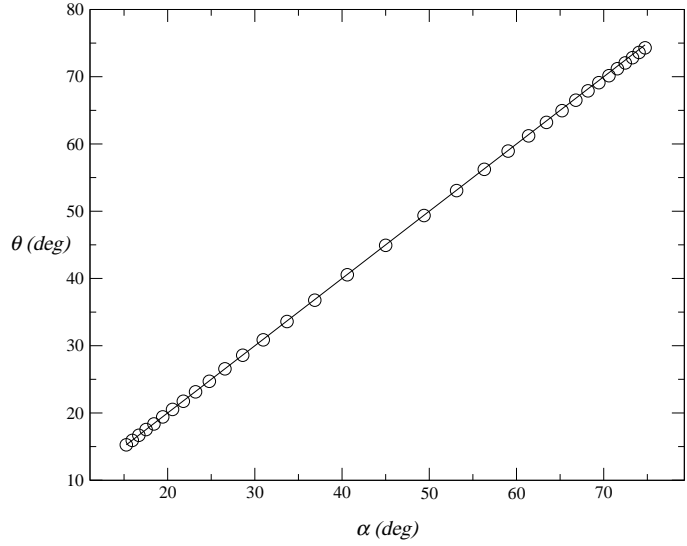}
\end{center}
\caption{Plot of the contact angle at the filling transition, $\theta$, 
against the wedge angle, $\alpha$. The error bars on $\theta$ lie within
the circles. The continuous line corresponds to the theoretical prediction
\Eref{thermo}.\label{fig3}}
\end{figure}

We have numerically minimised a discretized version of the continuum
Landau free-energy functional in a number of finite-size geometries with
boundary conditions chosen to mimic the bulk as closely as possible. We
considered a double wedge geometry with opposing surface magnetizations on the
top and bottom wedges and also a capped wedge geometry (see \Fref{fig2}) with 
both free and fixed boundary conditions at $z=L_1$. The results for the 
phase boundary and logarithmic growth of the filling layer are the same in all 
of these geometries. After testing our algorithm on a number of standard one- 
and two-dimensional problems including the Laplace and Poisson equations
we first applied the minimization scheme to the Landau free-energy 
functional with $\alpha=\pi/4$. We found that, as the system was
made larger and the discretization finer, our results stabilised and the 
phase boundary could be made arbitrarily close to the theoretical prediction. 
The smallest system with the roughest discretization that produced a clear
logarithmic growth and a transition value of $\theta$ within $0.1$ degrees of
$\alpha$ was $L_1=L_2\approx 30 \xi_b$ and a distance between the grid points of
approximately $\xi_b/2$. Taking this as a starting point we increased the 
number of points on the grid and reduced the grid spacing in the vertical 
direction, producing a more open wedge of the same depth, with the same 
discretization in the horizontal direction and a finer discretization in 
the vertical. The number of points and the vertical scale were adjusted to 
produce the various wedge angles. A similar procedure was used to produce 
acute wedges, but here the horizontal distance $L_2$ was kept constant 
and the wedge made gradually deeper. This approach allowed us study filling 
in wedges with a fairly wide  range of tilt angles, up to a maximum of 
about $70$ degrees, in a reasonable amount of CPU time. The effect of the 
variable discretization can be seen in \Fref{fig3}. For the most open wedges, 
$\theta$ and $\alpha$ agree to four significant figures as the fine 
discretization in the direction of the film growth, whilst for the very 
acute wedges $\theta$ and $\alpha$ may differ by as much as $0.5$ degrees.
\begin{figure}
\epsfxsize=9cm
\begin{center}
\epsfbox{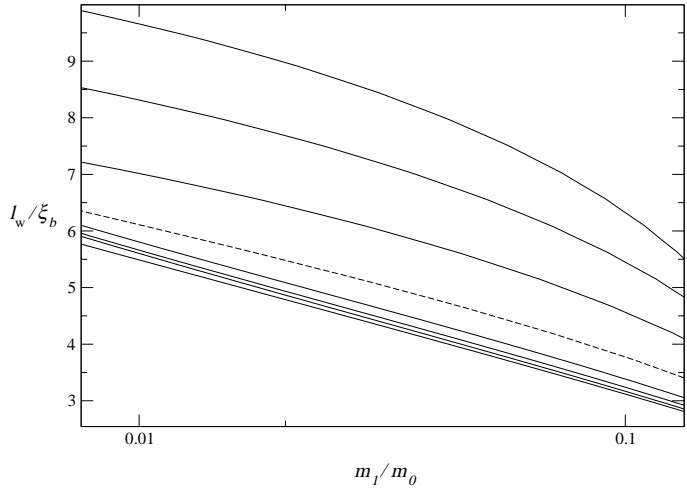}
\end{center}
\caption{Plot of the reduced midpoint interface height, $l_w/\xi_b$, against
the reduced surface magnetisation, $m_1/m_0$, for a range of $\alpha$ between
approximately $15^\circ$ (bottom) and $75^\circ$ (top). The dashed line
corresponds to $\alpha=45^\circ$.\label{fig4}}
\end{figure}

The results of our study are presented in \Fref{fig3} and \Fref{fig4}. 
In \Fref{fig3} we present data for the numerically determined phase boundary 
for tilt angles between $15$ and $75$ degrees which is in excellent agreement 
with the theoretical prediction $\theta=\alpha$. We have also studied the 
meniscus profile defined as the surface of isomagnetization $m=0$ 
(see \Fref{fig4}). For wedges with tilt angles $\alpha\le\pi/4$ there is 
clear evidence for the logarithmic growth of the mid-point filling height 
\begin{equation}
\kappa l_w=A \ln(m_1^{fill}-m_1)+C
\label{fit}
\end{equation}
with a universal, angle independent constant $A=1.02\pm 0.04$ consistent with 
the shallow wedge limit $A=1$. The constant $C$ is non-universal as is the size
of the asymptotic critical regime which decreases with increasing wedge angle.
Nevertheless even for the most acute angles we consider there is no evidence that
the amplitude $A$ differs from unity. More precisely for all wedge angles where
there is clear logarithmic growth of the filling layer the amplitude of the
logarithmic growth is consistent with $A=1$. In other words, even away 
from the shallow wedge limit the film diverges as
\begin{equation}
\kappa l_w(\theta,\alpha)\sim -\ln (\theta-\alpha)
\label{Acutecov}
\end{equation}
and behaves, in the critical regime, precisely like $l_{\pi}(\theta-\alpha)$.
This observation is certainly aesthetically appealing. In 2D, all results 
point to non-classical covariance for both shallow and more acute wedges
as indicated by exact Ising studies for the right-angle corner and also a 
drumhead model calculation with arbitrary wedge angle $\alpha$. The Landau
theory numerics, coupled with the interfacial model studies for the shallow
wedge, show that this is also the case for classical covariance. 
\begin{figure}
\epsfxsize=9cm
\begin{center}
\epsfbox{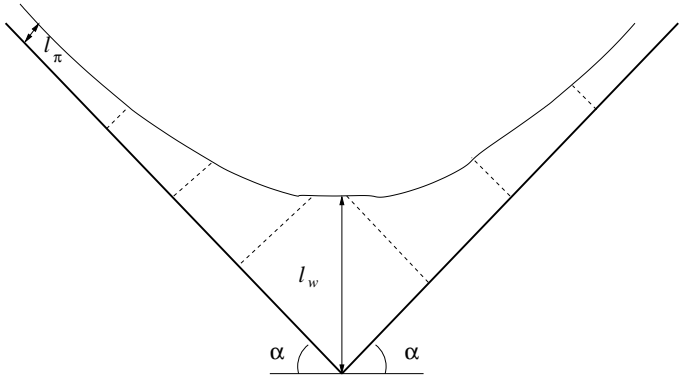}
\end{center}
\caption{Substrate-interface interaction paths (dashed lines) for the wedge 
filling transition in the model of Rejmer \etal \cite{Rejmer}.\label{fig5}}
\end{figure}

Given that the interfacial model (\ref{Heff}) is only valid for
shallow wedges its is natural to inquire whether one can explain the Landau
theory results using a generalised interfacial model valid for more acute
wedges. We finish this section by showing that a very reasonable, and for 
many purposes satisfactory, drumhead-like interfacial model proposed by Rejmer, 
Dietrich and Napi\'orkowski (RDN) \cite{Rejmer} does not fulfil this 
requirement.
The model is a straightforward generalization of the shallow wedge Hamiltonian
but is amended in two ways. First the square gradient term multiplying the
tension is replaced by the correct expression for the total area of the
liquid-vapour interface. This is familiar from older drumhead-like
capillary-wave models and ensures that the model recovers the correct filling
phase boundary $\theta=\alpha$. Secondly they propose replacing the relative 
vertical height from the wall to interface, appearing in the
generalised binding potential contribution, by the normal distance to the 
closest wall (see \Fref{fig5}). Thus the RDN model for the 2D/3D wedge is 
given by \cite{Rejmer}
\begin{equation}
H_{RDN}=\int d{\bf{x}} \left\{ \Sigma\left[\sqrt{1+\left(\nabla l \right)^2}-1
\right]+\sec\alpha W\left(\sec\alpha\left[l-\psi({\bf{x}}\right]\right)\right\}
\label{HRDN}
\end{equation}
where, as earlier, $l=l({\bf{x}})$ denotes the vertical height relative to the 
${\bf{x}}=0$ plane and $\psi({\bf{x}})=\tan\alpha\vert x\vert$. This assumption
is certainly valid far from the wedge bottom where the interface and wall are
parallel. One potentially unsatisfactory feature of it is that points on the
wall for which for distances $\vert x\vert <l_w \sin\alpha\cos\alpha$ do 
not contribute to the binding potential i.e. the interface is blind to the 
wedge bottom. The mean-field equation for the mid-point height, obtained 
by minimising the Hamiltonian and integrating once, is
\begin{equation}
\sigma_{lv}(\cos\theta-\cos\alpha)=W(l_w\cos\alpha)
\end{equation}
which is the analogue of the shallow wedge expression (\ref{EL}). In this way it 
is straightforward to see that the wedge filling transition occurs at
the correct thermodynamic phase boundary $\theta=\alpha$ and that the critical
exponents for continuous filling are independent of the wedge angle. 
Now consider the case of short-ranged forces modelled by the binding 
potential (\ref{Binding1}). According to the above drumhead model the 
asymptotic divergence of the mid-point height is given by
\begin{equation}
\kappa l_w\sim -\sec\alpha\ln(\theta-\alpha)
\label{wrong}
\end{equation}
and contains a geometrical amplitude factor $\sec\alpha$ which is not
consistent with the Landau theory numerics. In other words the RDN model does not 
give the correct, wedge-covariant result (\ref{Acutecov}).

The failure of the RDN model to account for classical wedge
covariance in more acute wedges is a surprising result. The model is certainly
plausible and would appear to be the simplest possible generalization of the
shallow wedge model. Indeed when considered beyond mean-field theory using
transfer matrix methods, the RDN model does predict non-classical wedge 
covariance for 2D systems with strictly short-ranged forces! We emphasise that
there is no mystery here; non-classical wedge covariance is a
fluctuation-induced phenomenon for which the precise form of the binding
potential (determining the way in which short-range forces are modelled) is
irrelevant. In the presence of large fluctuations with $l_w\sim\xi_{\perp}$ 
all points on the interface can feel the influence of the wall through 
collisions with it. Thus in the 2D FFL regime it is sufficient to use a 
binding potential with a simple square well shape representing pure contact 
forces. This contrasts with classical wedge covariance which reflects the 
precise form of the underlying interfacial Hamiltonian model. At mean-field 
level there are no fluctuation effects that take the interface to the wall 
and even for short-ranged forces the manner in which one models the large 
distance exponential tail of the binding potential is crucial. Clearly the 
assumption that the wall-interface interaction occurs via the normal 
distance to points on the closest wall is incorrect. The absence of a 
$\sec\alpha$ prefactor in the Landau numerics (\ref{Acutecov}) indicate 
that the correct measure of this is more akin to an effective local, vertical 
interaction similar to the shallow wedge model. This observation may well 
have ramifications for the construction of interfacial models of 3D wetting 
and filling in systems with short-ranged forces.

\section{3D wedge filling and the interfacial wandering exponent
\label{wandering}}

\subsection{Interfacial wandering and criticality at complete wetting 
\label{wandering1}}

In the treatment of fluctuation effects at 3D  wedge filling presented  
in the introduction there appears to be little connection between the 
values of the critical exponents and the interfacial wandering exponent 
$\zeta(d)$ defined for a planar-like fluid interface. The value of the 
wandering exponent $\zeta(d)$ is crucial for discussions of depinning as 
well as complete and critical wetting transitions. Interfacial fluctuation 
effects at such transitions are isotropic in the $d-1$ dimensions parallel 
to the interface and the wandering exponent describes the scaling relation 
between the roughness $\xi_{\perp}$ and transverse or parallel correlation 
length $\xi_{\parallel}$ \cite{Fisher}:
\begin{equation}
 \xi_{\perp}\sim\xi_{\parallel}^{\zeta(d)}
\label{zetad}
\end{equation}
In general the value of $\zeta(d)$ depends on the dimensionality of space and 
the qualitative type of disorder. For pure systems, thermal fluctuations lead
to interfacial roughness for $d\le 3$ and we have the well-known result
\cite{Fisher}
\begin{equation}
\zeta(d)=\frac{3-d}{2}
\label{thermalzeta}
\end{equation}
where for the marginal case, the identification $\zeta(3)=0$ corresponds to
$\xi_{\perp}\sim \sqrt{\ln\xi_{\parallel}}$. The lower critical dimension for
phase separation is $d_L=1$ at which $\zeta(d)=1$. The wandering exponent is
altered in the presence of quenched impurities and it is usual to distinguish
between random-field and random-bond disorder. Random fields induce interfacial
roughness for $d<5$ and general scaling arguments lead to the result
$\zeta(d)=(5-d)/3$ \cite{Fisher,Grinstein,Villain} implying a lower critical 
dimension at $d_L=2$. For random bonds on the other hand the explicit 
dimension dependence of $\zeta(d)$ is not known. Near the upper marginal 
dimension, $d=5$, approximate functional renormalization group calculations 
of an interfacial model lead to the linear relation \cite{Fisher,DFisher}
\begin{equation}
\zeta(d)\approx 0.2083(5-d)
\label{RG}
\end{equation}
This is believed to be accurate even far from the marginal dimension since for
$d=2$ it is close to the known exact result $\zeta(2)=2/3$ \cite{Huse}. 
In three dimensions numerical results suggest $\zeta(3)\approx 0.44$ 
\cite{Halpin-Healy}, slightly higher than the value predicted by the linear relation 
(\ref{RG}). The lower critical dimension for interfacial wandering induced 
by random-bond disorder is $d_L=5/3$ \cite{Huse2}, below which
they are unstable with respect to thermal fluctuations.

The wandering exponent explicitly determines the fluctuation-dominated
values of the critical exponents at the complete wetting transition. Complete
wetting is pertinent to planar wall-vapour interfaces (say) with vanishing
contact angle and refers to the divergence of the mean wetting layer (of
liquid) thickness $l$, roughness $\xi_{\perp}$ and transverse correlation
length $\xi_{\parallel}$ as the bulk two-phase coexistence is approached.
Writing the bulk ordering field $h\propto (\mu_{sat}(T)-\mu)$ we define 
critical exponents according to the power laws
\begin{equation}
l\sim h^{-\beta_s^{co}}\quad,\quad \xi_{\perp}\sim h^{-\nu_{\perp}^{co}}\quad,
\quad\xi_{\parallel}
\sim h^{-\nu_{\parallel}^{co}}
\label{completeexponents}
\end{equation}
The general theory of complete wetting is rather fully developed and can be
understood using effective interfacial Hamiltonian models \cite{Lipowsky4,
Lipowsky5}. The usual starting point for the discussion of fluctuation regimes 
is the continuum model (\ref{Heff}) with a binding potential 
\begin{equation}
W(l)=hl+al^{-p}
\label{combinding}
\end{equation}
where the Hamaker constant $a$ is positive. Transfer matrix and renormalization
group studies show that the critical behaviour falls in two scaling 
regimes \cite{Lipowsky4}. The values of the critical exponents in these 
regimes can also be understood using a simple effective potential picture. 
As noted by Lipowsky and Fisher \cite{Lipowsky5} the form of the bending 
energy in the interfacial model suggests that fluctuation effects can be 
accounted for by minimising an effective potential
\begin{equation}
W_{eff}(l)=hl+al^{-p}+b l^{-\tau}
\label{effcom}
\end{equation}
where $\tau=2(1/\zeta(d)-1)$. The complete wetting transition is therefore
fluctuation-dominated, belonging to the so-called weak-fluctuation (WFL)
regime, for $\tau<p$ and leads to the critical exponent identification
\begin{equation}
\beta_s^{co}=\frac {\zeta(d)}{2-\zeta(d)}
\label{betazeta}
\end{equation}
with $\beta_s^{co}=\nu_{\perp}^{co}=\zeta(d)\nu_{\parallel}^{co}$. This 
approach together with the final exponent identification will be important 
in our discussion of filling in ordered and disordered systems.

\subsection{From complete wetting to filling\label{wandering2}}

\begin{figure}
\epsfxsize=9cm
\begin{center}
\epsfbox{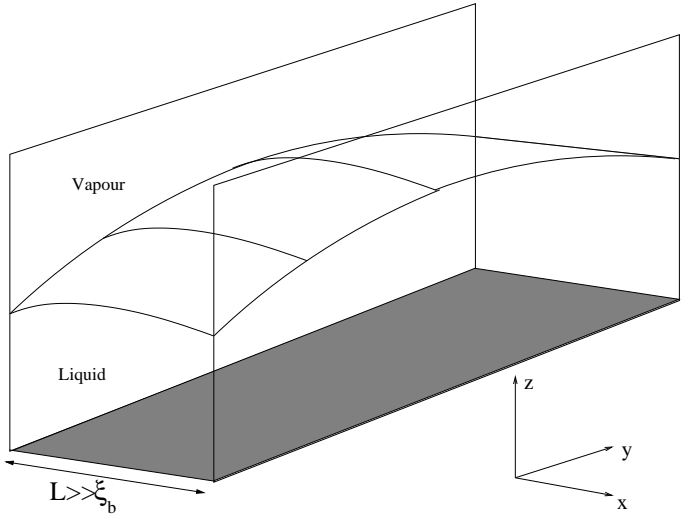}
\end{center}
\caption{The vapour-liquid interface at complete wetting in a constrained 
geometry across the $x-$axis. See text for explanation.\label{fig6}}
\end{figure}
Complete wetting and 3D filling share a number of common properties.
Both transitions exhibit two fluctuation regimes reflecting the fact that
there is always a relevant thermodynamic contribution to the appropriate
effective potential. The difference in the values of the fluctuation repulsion
exponents $\tau_w$ and $\tau$ can be traced directly to the
extra lengthscale dependence in the effective stiffness at filling. A relation
between these exponents can therefore be found if we understand how an 
additional lengthscale in the stiffness alters the fluctuation-induced 
repulsion. To this end consider a complete wetting transition occurring in 
a finite-size three-dimensional geometry with short-ranged forces 
(see \Fref{fig6}). The system is of infinite extent in the $y$ direction 
(say) but of finite width $L$ much bigger than the bulk correlation length. 
We may suppose that periodic boundary conditions apply across the system. 
In the presence of a finite bulk field $h$ the mean interface height is 
finite but diverges as $h\to 0$ with the growth of the film thickness falling 
into two regimes. If the width $L$ is larger than the transverse correlation 
length $\xi_{\parallel}\sim h^{-\nu_{\parallel}(3)}$ of a three-dimensional 
complete wetting layer of infinite extent, then the film grows with 3D-like 
critical exponents. If on the other hand $L\ll\xi_{\parallel}$ the growth of 
the wetting layer thickness must be characterised by critical exponents 
pertinent to 2D complete wetting. Standard finite-size scaling ideas imply 
that the crossover between these regimes should emerge from the scaling 
hypothesis
\begin{equation}
l(L,h)=h^{-\beta_s^{co}(3)}\Lambda (Lh^{\nu_{\parallel}(3)})
\label{FS}
\end{equation}
where $\Lambda(x)$ is an appropriate scaling function. In the limit
$x\to\infty$ the scaling function must approach a constant consistent with the
bulk 3D limit. On the other hand as the argument vanishes we must impose
\begin{equation}
\Lambda(x)\sim x^{\frac{ \beta_s^{co}(3)-\beta_s^{co}(2)}{\nu_{\parallel}(3)}}
\label{Lmbda}
\end{equation}
in order to recover the correct 2D limit. Thus the asymptotic divergence of $l$
as $h\to 0$ contains the finite-size dependence
\begin{equation}
l\sim h^{-\beta_s^{co}(2)} 
L^{\frac{ \beta_s^{co}(3)-\beta_s^{co}(2)}{\nu_{\parallel}(3)}}
\label{lLasymp}
\end{equation}
This asymptotic divergence should also be understandable using a suitably
modified effective potential measuring the free energy per unit length of the
finite-size system. As $h\to 0$ the interface behaves as a 2D-like complete
wetting layer in an effective bulk field $h_{eff}\propto hL$ and with modified
stiffness $\Sigma_{eff}\propto \Sigma L$. The divergence of $l$ should
therefore follow from the minimization of an effective 2D-like complete wetting
potential
\begin{equation}
W_{eff}(h,L)=(hL)l+AL^{-\phi}l^{-\tau(2)}
\label{WLH}
\end{equation}
where the exponent $\phi$ allows for the influence of the extra lengthscale
dependence in the effective stiffness. In order to recover the correct
finite-size result (\ref{lLasymp}) we must identify
\begin{equation}
\phi=\frac{\beta_s^{co}(2)-\beta_s^{co}(3)}{\beta_s^{co}(2)\nu_{\parallel}(3)}+\frac{1}{\beta_s^{co}(2)}-2
\label{phi}
\end{equation}

This exponent is key to understanding the connection between
interfacial wandering and 3D filling since it accounts for the correction to
the standard fluctuation induced repulsion at wetting when the effective
stiffness in the $y-$direction contains an additional length 
$\Sigma_{eff}=\Sigma L$. This is also the case at filling where the 
effective stiffness along the wedge is proportional to the interfacial 
height $l_w$. Simple power counting implies that the fluctuation-induced 
repulsion exponent for the wedge is
\begin{equation}
\tau_w=\phi+\tau(2)
\label{tauphi}
\end{equation}
which reduces to
\begin{equation}
\tau_w=\frac{2(1-\zeta(3))}{\zeta(2)}-1
\label{taufinal}
\end{equation}
In this way we can explicitly relate all the 3D wedge filling exponents in the
FFL regime to the values of the interfacial wandering exponent for 2D and 3D 
planar interfaces. Thus $l_w\sim\xi_{\perp}\sim\xi_y^{\zeta_w}$ with wedge 
wandering exponent
\begin{equation}
\zeta_w=\frac{\zeta(2)}{(1+\zeta(2)-\zeta(3))}
\label{zetafinal}
\end{equation}
which is valid provided both the two- and three-dimensional interfaces are well-defined and rough. That is the dimensionality must be such that $d\le d_u$ and
$d-1>d_L$. Similarly, and subject to the same provisos, the critical exponent 
for the interfacial height reduces to
\begin{equation}
\beta_w=\frac{\zeta(2)}{2(1-\zeta(3))}
\label{betafinal}
\end{equation}
which is the central result of this section. The interpretation and
implications of this result are discussed in detail below.

\subsection{Interpretation\label{wandering3}}

For pure systems the wandering exponent has values $\zeta(2)=1/2$ and
$\zeta(3)=0$ in two and three dimensions respectively. Consequently 
(\ref{zetafinal}) and (\ref{betafinal}) recover the effective Hamiltonian 
predictions $\zeta_w=1/3$ and $\beta_w=1/4$ discussed earlier. Further 
support for the connection between filling critical exponents and 
interfacial wandering arises when we generalise the above discussion 
to the $d$-dimensional wedge with $d-2$ dimensions along the direction of 
translational invariance. The analysis is unchanged except
that $\zeta(2)$ and $\zeta(3)$ are replaced by $\zeta(d-1)$ and $\zeta(d)$
respectively. Thus for a generalised wedge we predict
\begin{equation}
\beta_w=\frac{\zeta(d-1)}{2(1-\zeta(d))}
\label{genbetawd}
\end{equation}
provided that fluctuations are dominated by the $d-2$ dimensional breather
modes along the wedge. Substituting the thermal result $\zeta(d)=(3-d)/2$
recovers (\ref{betawd}) for $d<3$. For $d>3$ (\ref{genbetawd}) is no longer
valid since the $d$-dimensional interface is no longer rough. Notice that, in
the limit $d\to 2$, (\ref{genbetawd}) recovers the correct two-dimensional
filling result $\beta_w=1$ even though the breather-mode excitations are
effectively zero-dimensional, and that $\zeta(1)=1$.

For impure systems we are able to make some new predictions for
filling in random-bond and random-field systems. For random-bond systems we use
the values $\zeta(2)=2/3$ and $\zeta(3)=0.44$ to predict that in the FFL regime
\begin{equation}
\beta_w\approx 0.60\quad,\quad \nu_{\perp}\approx 0.60\quad,\quad 
\nu_y\approx 1.1
\label{RBFFL}
\end{equation}
which are valid for sufficiently short-ranged forces. Assuming that there are
only two fluctuation regimes and that there is a smooth cross-over from
mean-field to fluctuation-dominated exponents we anticipate that the above
predictions are valid provided the power law for the intermolecular forces
$p>1.7$. Thus dispersion-like (van der Waals) forces are irrelevant in the
renormalization group sense and belong to the universality class of systems
with short-ranged forces. Note that the value of the fluctuation repulsion
exponent $\tau_w$ is positive for both pure and random-bond systems. 

The case of 3D filling in systems with random-field disorder 
is less clear-cut. Using the values $\zeta(2)=1$ and $\zeta(3)=2/3$ we 
are led to the predictions $\beta_w=\nu_{\perp}=3/2$ and $\nu_y=2$. However
some caution is required in the assessing these predictions since
the reliance on the marginal value $\zeta(2)=1$ is somewhat unsatisfactory. 
This situation is reminiscent of the two-dimensional limit of 
(\ref{genbetawd}) for filling with thermal disorder which the relies on 
the marginal value $\zeta(1)=1$. The fact that the breather-mode picture generates the correct two-dimensional result for this case leads
some support to the above predictions for 3D random-field filling. However the
situation is worse for random-field disorder since the fluctuation repulsion
exponent $\tau_w$ is \emph{negative} for $d<10/3$. It is likely therefore that
the exponent identification (\ref{betafinal}) is not appropriate for 3D wedge
filling with random fields. Here after we limit our discussion to thermal 
and random-bond disorder only.

The above results for the values of the critical exponents for 
fluctuation-dominated three-dimensional wedge filling in pure and impure 
(random-bond) systems complement the known result for two-dimensional filling
(\ref{FFL2D}). It is therefore interesting to ask how the three-dimensional 
result (\ref{betafinal}) becomes the two-dimensional result as we compactify 
the wedge and reduce the number of dimensions along it. For pure systems 
with only thermal disorder the situation seems straightforward since the 
result (\ref{genbetawd}) is valid for $2\le d \le 3$. As noted above the 
generalised breather-mode result smoothly recovers the correct numerical 
result $\beta_w=1$ in this limit. There is however something remarkable 
about this since equating the two expressions (\ref{FFL2D}) and 
(\ref{genbetawd}) implies $\zeta(2)=\zeta(1)/2$ yielding information 
about the allowed value of the wandering exponent. Noting that the lower 
critical dimension for pure systems is $d_L=1$ and that at this dimension 
$\zeta=1$ we may conclude that equality of the breather-mode and 
wedge-covariant results for the critical exponents necessitates that for
thermal fluctuations
\begin{equation}
\zeta(d_L+1)=\frac{1}{2}
\label{zeta=1/2}
\end{equation}
which is correct. Turning attention to random-bond systems however note that 
we are not permitted to extend the breather mode result down to $d=2$ since 
the expression (\ref{genbetawd}) is not valid when $d-1<d_L=5/3$ equivalent 
to $d<8/3$. Therefore we are not able to equate distinct expressions for 
the filling critical exponents in two dimensions and are unable to discuss 
the values of these exponents for $2<d<8/3$. However we are still 
free to discuss the properties of the filling critical exponents as 
$d\to 8/3$ from above. In particular, in view of the thermal result
(\ref{zeta=1/2}), it is natural to inquire what the value of the wandering 
exponent is for random-bond systems one dimension above the lower critical 
dimension. That is, what is the value of $\zeta(8/3)$? As discussed earlier 
a reliable estimate for the value of the wandering exponent is provided by 
the linear relation (\ref{RG}) which is only slightly below the exact result 
in two dimensions. At $d=8/3$ this yields $\zeta(8/3)\approx 0.49$ which we 
anticipate to be only slightly below the true result. This remarkable numerical 
coincidence leads us to conjecture that (\ref{zeta=1/2}) is also valid for 
random-bond systems so that
\begin{equation}
\zeta\left(\frac{8}{3}\right)=\frac{1}{2}
\label{conjecture}
\end{equation}
It is natural to ask whether this conjecture has any support
from other examples of fluctuating surfaces. Lipowsky has considered
interfacial wandering and unbinding within a generalised class of effective 
Hamiltonians with bending energy term of the forms \cite{Forgacs} 
\begin{equation}
H[l]=\frac{K}{2}\int d{\bf{k}} k^{2-\eta} \vert \tilde l(k)\vert^2 + \int
d{\bf{x}} W(l({\bf{x}}))
\label{Lip}
\end{equation}
written in terms of the Fourier components of the collective co-ordinate. 
A remarkable property of such models is that for $\eta<0$ the
fixed point behaviour (pertinent to short-ranged critical unbinding \cite{Lipowsky6}) is
controlled by a single parameter rather than by $\eta$ and $d$ 
separately. For our purposes we note only that the wandering exponent 
$\zeta(d)=(3-d-\eta)/2$ and that the value at which this is unity allows us to 
identify $d_L=1-\eta$. This immediately implies that  $\zeta(d_L+1)=1/2$ 
consistent with the known result for interfaces in pure systems and our 
conjecture for interfaces subject to random-bond disorder. On the other
hand the result (\ref{zeta=1/2}) is not universally valid since for interfaces 
in random fields $\zeta=(5-d)/3$ implying $\zeta(d_L+1)=2/3$. Further work 
is required to substantiate the conjecture (\ref{conjecture}) for random-bond 
disorder.

\section{Tilt and torsional mode fluctuations at wedge filling
\label{tilttorsional}}

In this final section we return to the problem of wedge filling in
pure systems and reconsider the transfer matrix analysis of the wedge
Hamiltonian (\ref{Hw}). Bednorz and Napi\'orkowski \cite{Bednorz} have pointed 
out that some ambiguities in the construction of the infinitesimal transfer 
matrix arise because the bending energy term is proportional to the 
interfacial height. To avoid this they propose that the fluctuating field or 
order parameter appearing in the partition function functional integration is 
not $l_0(y)$ but rather $z(y)=l_0(y)^{3/2}$. The advantage of choosing this 
field is clear since the bending energy term is a simple Gaussian and one can 
map the problem immediately onto Euclidean quantum mechanics. Here we point out
that this choice of order parameter emerges naturally when we consider the role 
of tilt and torsional fluctuations coupled to breather-mode excitations.
\begin{figure}
\epsfxsize=9cm
\begin{center}
\epsfbox{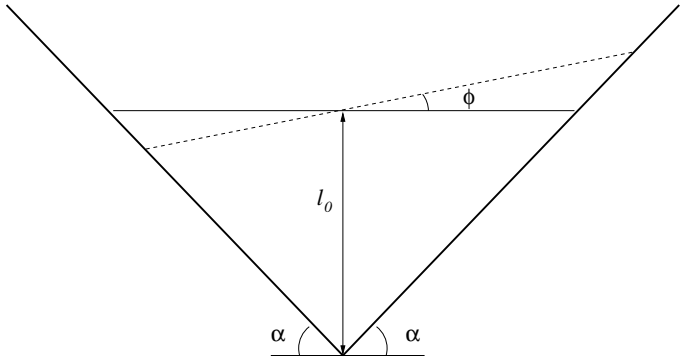}
\end{center}
\caption{Schematic picture of the tilt fluctuations of the filled region
in a section of a 3D wedge.
\label{fig7}}
\end{figure}

Using the wedge Hamiltonian we wish to evaluate the partition function
\begin{equation}
Z(l_1,l_2;L)=\int Dl_0 e^{-\beta H_{w}[l_0]}
\label{partition}
\end{equation}
for a wedge of length $L$ with fixed interfacial heights $l_1=l(0)$ and
$l_2=l(L)$ at the end points. The measure $Dl_0$ denotes some suitable 
integration over the allowed values of the collective co-ordinate $l(y)$ 
and will be discussed further below. The wedge model (\ref{Hw}) was 
constructed on the assumption that the only fluctuations that are relevant 
for filling correspond to local translations in the height of the filled 
region. Let us generalise this and allow for small, local tilts in the flat, 
filled region of the interface, that mimic the effect of the intrinsic
capillary fluctuations around the constrained profile. That is we consider 
a restricted class of interfacial configurations, that within the filled 
region has the form
\begin{equation}
l(x,y)=l_0(y) +\phi(y) x \quad;\quad l(x,y)< \alpha\vert x\vert
\label{restricted}
\end{equation}
provided $l(x,y)< \alpha\vert x\vert$. This is shown schematically in 
\Fref{fig7}. The tilt angle is assumed to lie in some interval
$-\epsilon\alpha\le \phi\le \epsilon\alpha$ with $0<\epsilon<1$. Outside of the
filled region we assume that the interface remains within a microscopic
distance from the wall so that $l(x,y)\approx\alpha\vert x\vert$. 
Substitution into the capillary-wave model leads to a wedge Hamiltonian 
that allows for breather, tilt and torsional modes and is a functional of 
the two fields $l_0$ and $\phi$. It is straightforward to show that the 
breather-tilt-torsional (BTT) model has the form
\begin{eqnarray}
H_{BTT}[l_0,\phi]=\int dy\Bigg\{\frac{K_1}{2}l_0\left(\frac{ dl_0}{dy}
\right)^2+K_2l_0^2\phi\left(\frac{ dl_0}{dy}
\right)\left(\frac{ d\phi}{dy}\right)\nonumber\\
+\frac{K_3}{2}l_0^3\left(\frac{ d\phi}{dy}
\right)^2 +V(l_0,\phi)\Bigg\}
\label{BTT}
\end{eqnarray} 
where the generalised wedge potential is
\begin{equation}
V(l_0,\phi)=\frac{\Sigma}{\alpha}(\theta^2-\alpha^2+\phi^2)l_0+Cl_0^{1-p}
\label{Vphi}
\end{equation}
The bending energy coefficient for the breather mode $K_1=\frac{2\Sigma}
{\alpha}$ and is unchanged from the simpler model ({\ref{Hw}}). The 
coefficients appearing in the torsional terms are $K_2\propto \Sigma
\alpha^{-2}$ and $K_3\propto\Sigma \alpha^{-3}$ respectively. The precise 
values of these coefficients are not important in the critical regime. 
It is also convenient to re-write the wedge potential as
\begin{equation}
V(l_0,\phi)=tl_0+Cl_0^{1-p}+\frac{\Sigma}{\alpha}\phi^2l_0
\label{Vphi2}
\end{equation}
Notice that when we set the field $\phi(y)=0$ at all points along the wedge
the BTT Hamiltonian reduces to the wedge model (\ref{Hwz}). To assess the
relevance of tilt and torsional modes on the critical behaviour of the model we
consider a rescaling of the co-ordinates and fields similar to that discussed
earlier. With the additional tilt field we now consider the properties of the
model under the mapping
\begin{equation}
y\to y'=\frac{y}{b}\quad,\quad l_0\to l_0'=\frac{l_0}{b^{\zeta_w}}\quad,\quad
\phi\to\phi'=\phi b^{2\zeta_w}
\label{phirescaling1}
\end{equation}
where, as before, the wedge wandering exponent $\zeta=1/3$. Under this
rescaling the coefficient of the breather-mode bending coefficient $K_1$
{\it{and}} the term $\phi^2 l_0$ appearing in the wedge potential remain
invariant. The linear temperature-like scaling field $t$ and Hamaker constant
$C$ rescale as earlier (see (\ref{simplerescaling2})) whilst the new tilt and
torsional bending energy coefficients renormalise to
\begin{equation}
K_2'=b^{-4/3}K_2\quad,\quad K_3'=b^{-4/3}K_3
\label{Krescaling}
\end{equation}
and are therefore irrelevant for all ranges of intermolecular force. 
This means that they do not affect the critical behaviour and that the 
BTT model can be simplified to
\begin{equation}
H_{BTT}[l_0,\phi]=\int dy \left\{ \frac{\Sigma l_0}{\alpha} 
\left(\frac{ dl_0}{dy} \right)^2
+V(l_0,\phi) \right\}
\label{Hwphi}
\end{equation}
The tilt field $\phi$ is therefore non-interacting but still coupled to the
breather-mode excitations. As the interface unbinds from the wedge bottom the
tilt fluctuations become increasingly massive and are suppressed. An important
consequence of this is that when we integrate out the tilt field the 
particular choice of the parameter $\epsilon$ becomes 
unimportant in the scaling limit.

The final ingredient in the construction of the model is the nature of
the measure $Dl_0D\phi$ appearing in the functional integration over the
Boltzmann weight $e^{-\beta H_{BTT}[l,\phi]}$, where $\beta\equiv 1/k_B T$. 
In reducing the dimensionality of the collective co-ordinate $l(x,y)\to 
(l(y),\phi(y))$ some care is required in defining the new measure since we 
must impose the correct normalization conditions. This is most easily 
illustrated for 2D wedge filling in which there is no $y$-direction and 
the choice of measure $Dl_0D\phi$ relates only to one $l$ and one $\phi$ 
variable. For this case the Boltzmann weight for a tilt configuration is 
simply $e^{-\beta V(l,\phi)}$ and determines the \emph{un-normalised} 
probability $P(l_0,\phi)$ for finding an interface with tilt angle $\phi$ 
and mid-point height $l_0$. Integrating over the allowed values of $\phi$ 
determines the mid-point height PDF $P(l_0)$. The properties 
of this PDF are known from exact calculations. In particular for
short-ranged forces (belonging to the filling fluctuation regime) the function
has the simple exponential form $P(l_0)\propto e^{-2\beta\Sigma(\theta-
\alpha)l_0}$ in the scaling limit. Thus we require that as $\theta\to\alpha$
\begin{equation}
e^{-2\beta\Sigma(\theta-\alpha)l_0}\sim \int_{-\epsilon\alpha}^{\epsilon\alpha} 
Dl_0d\phi e^{-\beta\Sigma(\theta^2-\alpha^2+\phi^2)l_0/\alpha}
\label{2Dmeasure}
\end{equation}
which necessitates that we interpret the 2D measure $Dl_0\propto \sqrt{l_0} 
dl_0$. In other words the correctly normalised PDF $P(l_0,\phi)\propto 
\sqrt{l_0}e^{-\beta V(l,\phi)}$. An equivalent argument is that after 
integrating over the tilt fluctuations the resulting PDF $P(l_0)$ must have 
the correct short-distance expansion as the interface approaches the wall. For 
2D systems with short-ranged forces the PDF has the short-distance expansion 
$P(l_0)\sim l_0^{\gamma}$ with $\gamma=1/\beta_w-1$. Recall that for pure 
systems $\beta_w=1$ (in 2D) implying that the two-field PDF $P(l_0,\phi)$ 
must contain a $\sqrt {l_0}$ prefactor in-order to preserve thermodynamic 
consistency. This argument is supported by explicit results for the 
2D wedge filling for contact binding potentials (see \ref{appendix}).

A similar line of reasoning applies in higher dimensions. To determine the form
of the 3D measure we first focus on the filling transition occurring in a cone
geometry. The reason for this is that the breather-mode excitations are
essentially zero-dimensional and we only have one height variable $l_c$
measuring the height of the interface above the cone vertex. Up to a
normalization factor the probability of finding the interface at height 
$l_0$ with tilt $\phi$ is determined by the Boltzmann weight $e^{-\beta V_c}$ 
where $V_c$ is the free-energy cost of the configuration. The leading order 
term in the cone binding potential contains a term proportional to the area 
$l_0^2$ of the filled region. The prefactor of this is $\beta\Sigma(\theta^2-
\alpha^2+\phi^2)$ similar to the 2D wedge and integrating over the tilt modes 
(and a trivial rotational degree of freedom) determines the mid-point height
PDF $P_c(l_c)$. In order to preserve the simple Gaussian form of this function
we must interpret the 3D measure $Dl_c\propto l_cdl_c$. For the 3D wedge 
we discretise the system in the $y-$direction and replace the fields $\{l(y)\}$
and $\{\phi(y)\}$ by sets of continuous variables $l_j$ and $\phi_j$. The 
wedge measure must be proportional to a product of cone measures and thus we 
interpret the functional integral for the BTT Hamiltonian as 
\begin{equation}
\int Dl_0 D\phi\propto \int\int\int\prod_{j}l_0(j)dl_0(j)d\phi_j
\label{wedgemeasure}
\end{equation}
On integrating over each tilt angle we recover the wedge model
(\ref{Hw}) and identify the functional integral appearing in (\ref{partition})
as the continuum limit of
\begin{equation}
\int\int\ldots\int\prod_{j}l_0(j)^{1/2}dl_0(j)
\label{finalmeasure}
\end{equation}
In terms of the variable $z_j=l_0(j)^{3/2}$ the functional integral therefore
becomes 
\begin{equation}
\int\int\ldots\int\prod_{j}dz(j)
\label{finalmeasure2}
\end{equation}
and has a well-defined infinitesmal transfer matrix limit. In terms of the new
variable the continuum wedge Hamiltonian is
\begin{equation}
\tilde H_w[z]=\int dy \left\{ \frac{4\Sigma}{9\alpha} \left(\frac{ dz}{dy} \right)^2
+\tilde V(z) \right\}
\label{Hwz}
\end{equation}
where $\tilde V(z)\equiv V(z^{2/3},0)$. The rest of the analysis is identical
to that described by Bednorz and Napi\'orkowski \cite{Bednorz} and the problem 
can be mapped immediately on to quantum mechanics. Thus the partition 
function may be written as the spectral expansion
\begin{equation}
Z(l_1,l_2;L)=\sum_{n}\psi_n(l_1)^*\psi_n(l_2)e^{-\beta E_n L}
\label{spectral}
\end{equation}
where the eigenfunctions and eigenvalues satisfy the Schr\"odinger equation
\begin{equation}
-\frac{9\alpha}{16\beta^2\Sigma}\psi_n(z)''+\tilde V(z)\psi_n(z)=E_n\psi_n(z)
\label{Shrodinger}
\end{equation}
with boundary conditions $\psi(0)=\psi(\infty)=0$. Hence the PDF for
finding the interface at height $l_0$ can be identified with
$P_w(l_0)=\vert\psi_0(l^{3/2})\vert ^2$ whilst the wedge free energy and
correlation length follow from $f_w=E_0$ and $\xi_y^{-1}=\beta(E_1-E_0)$ 
respectively. For the choice of wedge potential (\ref{wedgepotential}) it is
straightforward to deduce the existence of two fluctuation regimes with the
critical exponents discussed earlier.

\begin{figure}
\epsfxsize=10cm
\begin{center}
\epsfbox{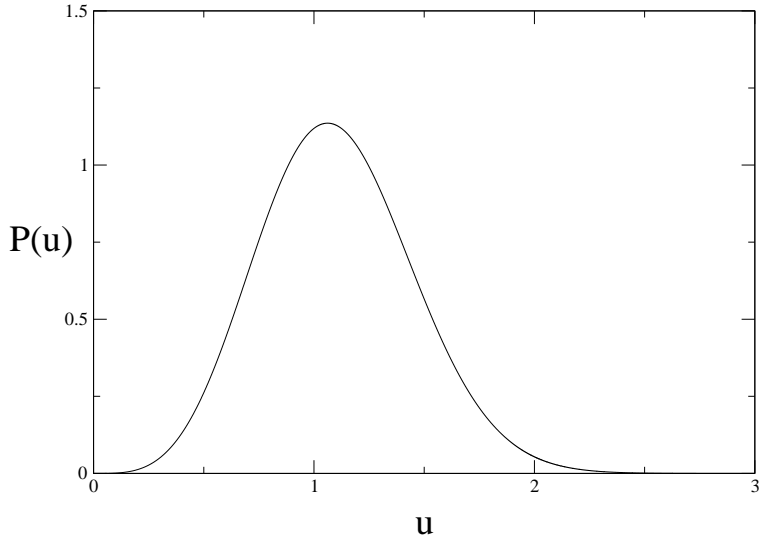}
\end{center}
\caption{Rescaled probability distribution function for the
interfacial height above the wedge bottom found from the breather
mode model. Integration of the PDF determines the density profile in the 
wedge.
\label{fig8}}
\end{figure}

Our final remarks concern the form of the interfacial height PDF 
$P(l_0)$ in the FFL regime. The intermolecular forces are irrelevant for
$p>4$ and in the limit $\theta\to\alpha$ the scaling properties of the PDF
follow from solution of the differential equation
\begin{equation}
-\frac{9\alpha}{16\beta^2\Sigma}\psi_0(z)''+2\Sigma(\theta-\alpha)
z^{\frac{2}{3}}\psi_0(z)=E_0\psi_0(z)
\label{Schrodinger2}
\end{equation}
It follows immediately that the wave-function and hence PDF are scaling 
functions of the dimensionless variable $u=\sqrt {\beta\Sigma} 
(\theta/\alpha-1)^{1/4}l_0$ and we can write
\begin{equation}
P(l_0)=\sqrt {\beta\Sigma} \left(\frac{\theta}{\alpha}-1\right)^{1/4}g
(\sqrt {\beta\Sigma} (\theta/\alpha-1)^{1/4}l_0)
\label{PDFscaling}
\end{equation}
where $g(u)$ is a suitably normalised scaling function such that $\int
du g(u)=1$. The scaling of the PDF is, of course, to be anticipated since
in the FFL both $l_w$ and $\xi_{\perp}$ diverge with the same critical
exponent. We are not aware of a closed form solution to the differential
equation in terms of elementary functions. However both the asymptotic 
short-distance and large-distance behaviours agree with thermodynamic and 
sum-rule requirements. First consider the large distance limit corresponding 
to $u\gg 1$. From solution of (\ref{Schrodinger2}) it is straightforward to 
show that the PDF decays like
\begin{equation}
P(l_0)\sim \sqrt {\beta\Sigma}
(\theta/\alpha-1)^{1/4}e^{-2\beta\Sigma\sqrt{2(\theta/\alpha-1)}l_0^2}
\label{largedistance}
\end{equation}
up to terms of order $l_0$ in the argument. This is precisely in keeping with
macroscopic requirements. When the interface is far from the wall the PDF can
be identified with $e^{-\beta \Delta F}$ where $\Delta F$ is the free-energy
cost of a macroscopic configuration constrained to pass through the height 
$l_0$. In the wedge geometry this macroscopic configuration has the shape shown
in \Fref{fig7} and the free-energy cost is dominated by a thermodynamic 
area-like contribution. This can be readily calculated making use of Young's 
equation yielding $\Delta F=2\Sigma\sqrt{2(\theta-\alpha)/\alpha}l_0^2$. The 
short-distance expansion (SDE) of the scaling function $g(u)$ corresponding 
to the limit $u\ll 1$ is also consistent with known critical exponent 
relations. For 2D and 3D wedge filling the scaling function for the PDF has 
the short-distance behaviour
\begin{equation}
g\sim u^\gamma
\label{SDE1}
\end{equation} 
where the SDE critical exponent $\gamma=1/\beta_w-1$. This exponent relation
follows from rather general sum-rule and scaling arguments. Thus for 3D filling
in pure systems with $\beta_w=1/4$ we anticipate 
\begin{equation}
P(l_0)\sim (\beta \Sigma)^2 \left(\frac{\theta}{\alpha}-1\right) l_0^3
\label{SDE2}
\end{equation}
for $\sqrt {\beta \Sigma} (\theta/\alpha-1)^{1/4}l_0\ll 1$. This emerges 
naturally from solution to the Schr\"odinger equation since in terms of the $z$ 
variable $\psi(z)\sim z$ as $z\to 0$ and recall $P(l_0)=\vert\psi(l_0^{3/2})
\vert^2$. The large-distance and SDE properties of the PDF give strong 
support to the correctness of the one-dimensional wedge Hamiltonian model.
The numerically determined scaling form of the PDF found from solution to
the Schr\"odinger equation is shown in \Fref{fig8}.

\section{Conclusions \label{conclusions}}

In this paper we have addressed a number of issues pertaining to 3D
wedge filling transitions. We began with a discussion of wedge covariance
relations between filling and critical wetting and showed that the 
fluctuation-induced covariance observed for 2D filling has a mean-field
analogue. Similar to the 2D case classical covariance appears to be specific to
systems for which the critical wetting specific heat exponent $\alpha_s=0$.
Numerical studies of mean-field filling within a Landau theory show that
classical covariance is not limited to the shallow wedge limit. Importantly 
this provides a constraint on the possible form of effective Hamiltonian 
models of wedge filling and implies that a model suggested by Rejmer \etal 
\cite{Rejmer} is not consistent with more microscopic approaches. It is 
likely that in order capture the correct classical covariance for wetting 
and filling it will be necessary to construct non-local interfacial models. 
This will be discussed elsewhere.

In the second part of our paper we turned attention to fluctuation
effects at 3D wedge filling. Our central new result is that the critical 
exponents which characterise the fluctuation-dominated regime can be related 
to the values of the standard wandering exponent for \emph{planar}
interfaces in 2D and 3D (bulk) systems. The expressions for $\beta_w$ and the
wedge wandering exponent $\zeta_w$ expressed in terms of $\zeta(2)$ and
$\zeta(3)$ are rather general and allow the discussion of 
filling in systems with random-bond disorder. A surprising
feature emerging from our study concerns some intriguing numerical 
coincidences for the value of the bulk wandering exponent $\zeta(d)$ 
as one approaches the lower marginal dimension for wedge filling in pure 
and random-bond systems. In particular we are led to conjecture that for 
random-bond disorder $\zeta(8/3)=1/2$.

In the final part of our paper we addressed some queries that have
been highlighted concerning the pseudo-one-dimensional wedge Hamiltonian 
theory of 3D filling in pure systems. Bednorz and Napi\'orkowski \cite{Bednorz}
have suggested that these can be avoided if one supposes that the 
fluctuating field is not the interfacial height $l_0(y)$ above
the wedge bottom but $l_0(y)^{3/2}$. We showed that this choice indeed 
emerges naturally if one considers the coupling between tilt and torsional 
fluctuations described by a second field $\phi(y)$ and breather-mode 
excitations of $l_0(y)$. The form of the PDF for the interfacial height 
predicted by the wedge Hamiltonian model is shown to have the correct 
short-distance and large-distance behaviour dictated by macroscopic 
arguments and critical exponent relations. We believe this strongly supports
the internal consistency of the effective wedge Hamiltonian model. As mentioned
in the introduction the values of the critical exponents predicted by the 
wedge Hamiltonian are in very good agreement with Ising model simulation 
studies. A more stringent test of the theory would be to study the scaling 
of the interfacial height PDF. This can be readily extracted from Ising 
model studies since it corresponds to the derivative of the magnetization 
profile measured along the vertical above the wedge bottom. To do this 
quantitatively it would be necessary to generalise the present theory to 
study the finite-size scaling of the PDF and magnetization profile in the 
same geometry studied by Milchev \etal \cite{Milchev1,Milchev2}. This will 
be a topic of future work.

Finally we mention that ideally one would hope to circumvent the entire
wedge Hamiltonian theory and replace it with a more microscopic approach. In
particular it would be extremely useful if one could apply a renormalization
group analysis to derive the wedge filling critical exponents from the full 
capillary-wave model (\ref{Heff}). To do this one would have to identify a 
new fixed-point distinct from the Gaussian (or strong-fluctuation regime) 
fixed-point used to study wetting transitions. Such additional fixed
points for the capillary-wave model may exist if one generalises the spatial 
rescaling to account for the anisotropy of correlation lengths across and 
along the wedge. One would hope, of course, that such an approach would lead to
the same predictions for universal critical exponents and scaling functions
found using the effective one-dimensional wedge Hamiltonian. As well as being
an independent check on the validity of the wedge Hamiltonian model (in the
scaling limit) such an approach would also shed much light on the emergence of
an effective one-dimensional theory from a higher dimensional model.

\ack
The authors wish to thank Dr A. J. Wood for discussions concerning the
nature of wedge filling in disordered systems, and I-D Media AG, Berlin,
for the donation of a workstation. MJG and JMR-E gratefully acknowledge 
the EPSRC and Secretar\'{\i}a de Estado de Educaci\'on y Universidades 
(Spain), co-financed by the European Social Fund for financial support 
respectively.
\appendix
\section{Tilt mode fluctuations at 2D wedge filling\label{appendix}}

\begin{figure}
\epsfxsize=9cm
\begin{center}
\epsfbox{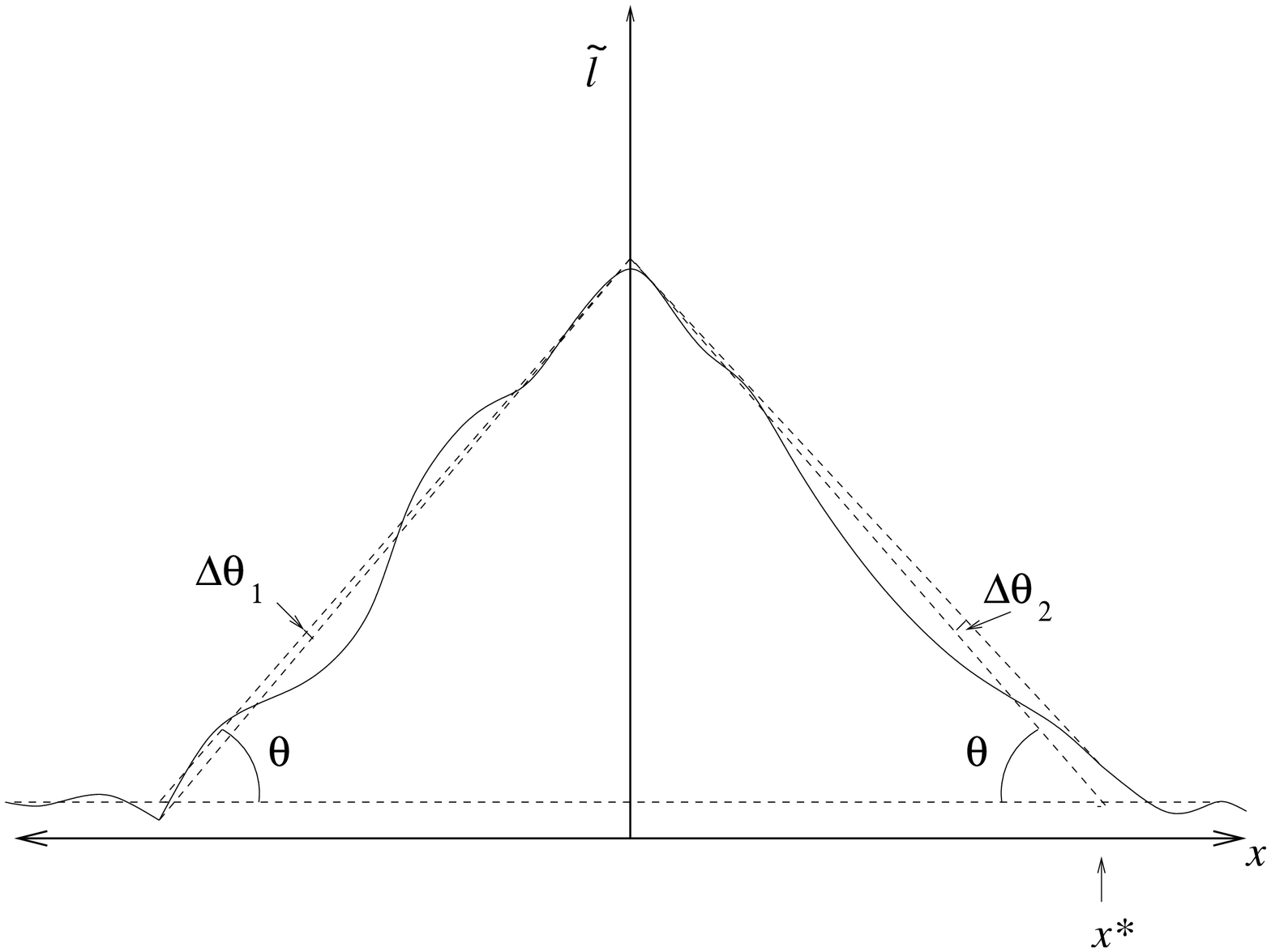}
\end{center}
\caption{Plot of the typical relative interfacial position $\tilde l\equiv l-
\alpha|x|$ close to the 2D wedge filling transition. $\Delta \theta_1$ and
$\Delta \theta_2$ define the tilt fluctuations of the interface at
$x^*$. See text for explanation.\label{fig9}}
\end{figure}

In this Appendix we provide further justification for the role played by tilt
fluctuations at 2D wedge filling making contact with exact results obtained
from interfacial Hamiltonian studies. We restrict attention to systems with
short-ranged forces corresponding to the FFL regime and set $k_B T=1$ for 
convenience. Consider the $3-$point probability distribution function 
$P_w(l_1,-x;l_0,0;l_2,x)$ determining the probability for finding the interface
at heights $l_1$, $l_0$ and $l_2$ at positions $-x$, $0$ and $x$, respectively.
Exact transfer-matrix results for a continuum interfacial model of filling in
shallow wedges show that $P_w(l_1,-x;l_0,0;l_2,x)$ decomposes into the
product \cite{Romero}
\begin{equation}
P_w(l_1,-x;l_0,0;l_2,x)=P_w(l_0)
P_\pi^c(\tilde{l}_1,x|l_0,0) P_\pi^c(\tilde{l}_2,x|l_0,0)
\label{3pPDF}
\end{equation} 
where $P_w(l_0)=2\Sigma(\theta-\alpha)\exp[-2\Sigma(\theta-\alpha)l_0]$ is 
the wedge midpoint $1-$point PDF, $\tilde{l}=l-\alpha x$ is the local relative
height to the wall (recall that $x$ is positive) and 
$P_\pi^c(l_2,x_2|l_1,x_0)$ is the planar conditional probability 
distribution function \cite{Romero}:
\begin{eqnarray}
P^c_\pi(l_2,x|l_1,0)=\sqrt{\frac{\Sigma}{2\pi x}}\textrm{e}^{-
\frac{\Sigma(l_2-l_1+\theta x)^2}{2x}}
+\textrm{e}^{- 2\Sigma\theta l_2}\Bigg[\sqrt{\frac{\Sigma}{2\pi x}}
\textrm{e}^{-\frac{\Sigma(l_1+l_2-\theta x)^2}{2x}}
\nonumber\\
+ \Sigma\theta \textrm{erfc}\left(\sqrt{\frac{ \Sigma}{2\pi x}}
(l_1+l_2-\theta x)\right)\Bigg]
\label{defconditional}
\end{eqnarray}
If the wedge midpoint height is much greater than the mean wetting height
for a planar substrate, i.e. $l_0\gg 1/2\Sigma \theta$, which is fulfilled
in the asymptotic critical regime and the position $x<l_0/\theta$, 
the second term in \Eref{defconditional} can be 
neglected and the conditional probability distribution function becomes a 
gaussian distribution of average $l_0-\theta x$ and dispersion $\sigma^2 =
x/\Sigma$. The interface will be affected by the presence of the substrate 
when $l_0-\theta x\sim \sqrt{x/\Sigma}$. This expression defines the effective
value of $x^*$ at which the interface is in contact with the substrate.
We consider the angle formed by the horizontal and the line that joins the 
interface position at the midpoint and the position (relative to the 
substrate) at $x^*$ (see \Fref{fig9}). The tilt angles $\Delta \theta_1$ and
$\Delta \theta_2$ are \emph{defined} as the departures of the previously 
defined angles with respect to the average value $\theta$. We stress 
that the interface is not considered stiff but is allowed to wander 
due to capillary-wave fluctuations. However, the length scale on which 
these fluctuations occur is much smaller than the one that characterizes 
the constrained profile. When $\Delta \theta_{1,2} \ll \theta \ll 1$, 
then $l_{1,2}-l_0+\theta x^*\approx x^* \Delta \theta_{1,2}$.

Under these assumptions, the PDF for a wedge midpoint position $l_0$ and
tilt angles $\Delta \theta_1$ and $\Delta \theta_2$ can be approximated as
\begin{equation}
P(l_0,\Delta \theta_1,\Delta \theta_2)\approx 2\Sigma (\theta-\alpha) 
e^{-2\Sigma (\theta-\alpha) l_0}\frac{\Sigma x^*}{2\pi} e^{-\frac{\Sigma x^*}{2}
\left[(\Delta\theta_1)^2+(\Delta\theta_2)^2\right]}  
\label{plwtheta12}
\end{equation} 
where the tilt angle ranges can be extended safely to the complete real axis.
We \emph{define} the tilt angles $\phi$ and $\phi'$ as $(\Delta \theta_1+
\Delta \theta_2)/2$ and $(\Delta \theta_1-\Delta \theta_2)/2$, respectively. 
Taking into account that in the scaling limit $x^*\approx l_0/
\theta \approx l_0/\alpha$, \Eref{plwtheta12} can be expressed as:
\begin{eqnarray}
P(l_0,\phi,\phi')&\approx& \frac{2 \Sigma^{3/2} \sqrt{l_0}}
{\sqrt{\pi \alpha}} (\theta-\alpha) e^{-\frac{\Sigma l_0}{\alpha}[2\alpha 
(\theta-\alpha)+\phi^2]}\nonumber\\
&\times& \sqrt{\frac{\Sigma l_0}{\pi \alpha}}
e^{-\frac{\Sigma l_0 (\phi')^2}{\alpha}}
\label{plwtheta12-2}
\end{eqnarray}
Integrating all the values of $\phi'$, we obtain:
\begin{eqnarray}
P(l_0,\phi)\approx \frac{2 \Sigma^{3/2} \sqrt{l_0}}
{\sqrt{\pi \alpha}} (\theta-\alpha) e^{-\frac{\Sigma l_0}{\alpha}[2\alpha 
(\theta-\alpha)+\phi^2]}\nonumber\\
\approx \frac{2 \Sigma^{3/2} \sqrt{l_0}}
{\sqrt{\pi \alpha}} (\theta-\alpha) e^{-\frac{\Sigma l_0}{\alpha}(
\theta^2-\alpha^2+\phi^2)}
\end{eqnarray}
in agreement with \Eref{2Dmeasure} and with the correct measure $Dl_0
\sim \sqrt{l_0} dl_0$. 

\end{document}